\documentclass[twocolumn,showpacs,preprintnumbers,amsmath,amssymb,prb]{revtex4-1}
\usepackage{soul}
\usepackage[bbgreekl]{mathbbol}
\usepackage{graphicx}
\usepackage{dcolumn}
\usepackage{bm}
\usepackage{mathtools}
\usepackage{hyperref}
\usepackage{color}
\usepackage[a4paper]{anysize}
\graphicspath{{fig/}{img/}}
\usepackage[usenames,dvipsnames,svgnames,table]{xcolor}

\newcommand{\E}{{\bf E}}

\renewcommand{\k}{{\bm{k}}}

\def\gsim{\lower.35em\hbox{$\stackrel{\textstyle>}{\textstyle\sim}$}}
\def\lsim{\lower.35em\hbox{$\stackrel{\textstyle<}{\textstyle\sim}$}}

\begin{document}
\title{Plasmonic Dirac Cone in Twisted Bilayer Graphene}
\author{Luis Brey}
\affiliation{Materials Science Factory, Instituto de Ciencia de Materiales de Madrid (CSIC), Cantoblanco, 28049 Madrid, Spain}
\author{T. Stauber}
\affiliation{Materials Science Factory, Instituto de Ciencia de Materiales de Madrid (CSIC), Cantoblanco, 28049 Madrid, Spain}
\author{T. Slipchenko}
\affiliation{Instituto de Ciencia de Materiales de Arag\'on and Departamento de  F\'{\i}sica de la Materia Condensada, CSIC-Universidad de Zaragoza, E-50009, Zaragoza, Spain}
\author{L. Mart\'{\i}n-Moreno}
\affiliation{Instituto de Ciencia de Materiales de Arag\'on and Departamento de  F\'{\i}sica de la Materia Condensada, CSIC-Universidad de Zaragoza, E-50009, Zaragoza, Spain}
\date{\today}
\begin{abstract}
We discuss plasmons of biased  twisted bilayer graphene when the Fermi level lies inside the gap. The collective excitations are  a network of chiral edge plasmons (CEP) entirely composed of
excitations in the topological electronic edge states (EES) that appear at the AB-BA interfaces.
The CEP form an hexagonal network with an unique   energy scale $\epsilon_p=\frac{e^2}{\epsilon_0\epsilon t_0}$ with $t_0$ the moir\'e lattice constant and $\epsilon$ the dielectric constant. 
From the dielectric matrix we obtain the plasmon spectra that has two main characteristics: (i) a diverging density of states at zero energy, and (ii) the presence of a plasmonic Dirac cone at $\hbar\omega\sim\epsilon_p/2$ with sound velocity $v_D=0.0075c$, which is formed by zigzag and armchair current oscillations. 
A  network model reveals that the antisymmetry of the plasmon bands implies that CEP scatter at the hexagon vertices 
maximally in the deflected
chiral outgoing directions, with a current ratio of 4/9 into each of the deflected directions and 1/9 into the forward one.
We show that scanning near-field microscopy should be able to observe the predicted plasmonic Dirac cone and its broken symmetry phases.
\end{abstract}
\maketitle

\noindent {\it Introduction.}
The study of graphene has brought to light many unexpected basic as well as applied physical properties.\cite{Guinea_2009,Katsnelson-book}
More surprises appear when two graphene layers are stacked and rotated one on top of the other, forming the so-called twisted bilayer graphene (TBG). When the rotation angle is large, the graphene layers are electronically decoupled,\cite{Lopes_2007,Morell_2011b} but at small twist angles the Fermi velocity of the carriers reduces considerably\cite{Shallcross_2010}. At some particular magic angle, the
electronic bands become 
almost flat around charge neutrality.\cite{Suarez_2010,Bistritzer2011,Brihuega_2012} In this 
regime, new and unpredicted electronic phases emerge.\cite{Cao:2018aa,Cao:2018ab,Yankowitz19,Sharpe19}
Collective modes in twisted structures have been studied and show several new features.\cite{Stauber:2013aa,tom14,Ni15,Stauber:2016aa,Hu17,Stauber18,Sunku:2018aa,Levitov19, hesp2019collective, novelli2020optical} In this Letter, however, we will discuss the collective excitations associated to the network of chiral EES that appear when an electric field is applied perpendicular to the sample and the chemical potential lies inside the gap\cite{McCann_2006,Martin_2008,Jaskolski:2018aa}. In this regime, plasmons can be described by a simple macroscopic model that is defined by a hexagonal lattice of alternating Hall conductivities, as shown in Fig. \ref{Figure1} (d). 

\begin{figure}[h!]
\includegraphics[width=8.cm]{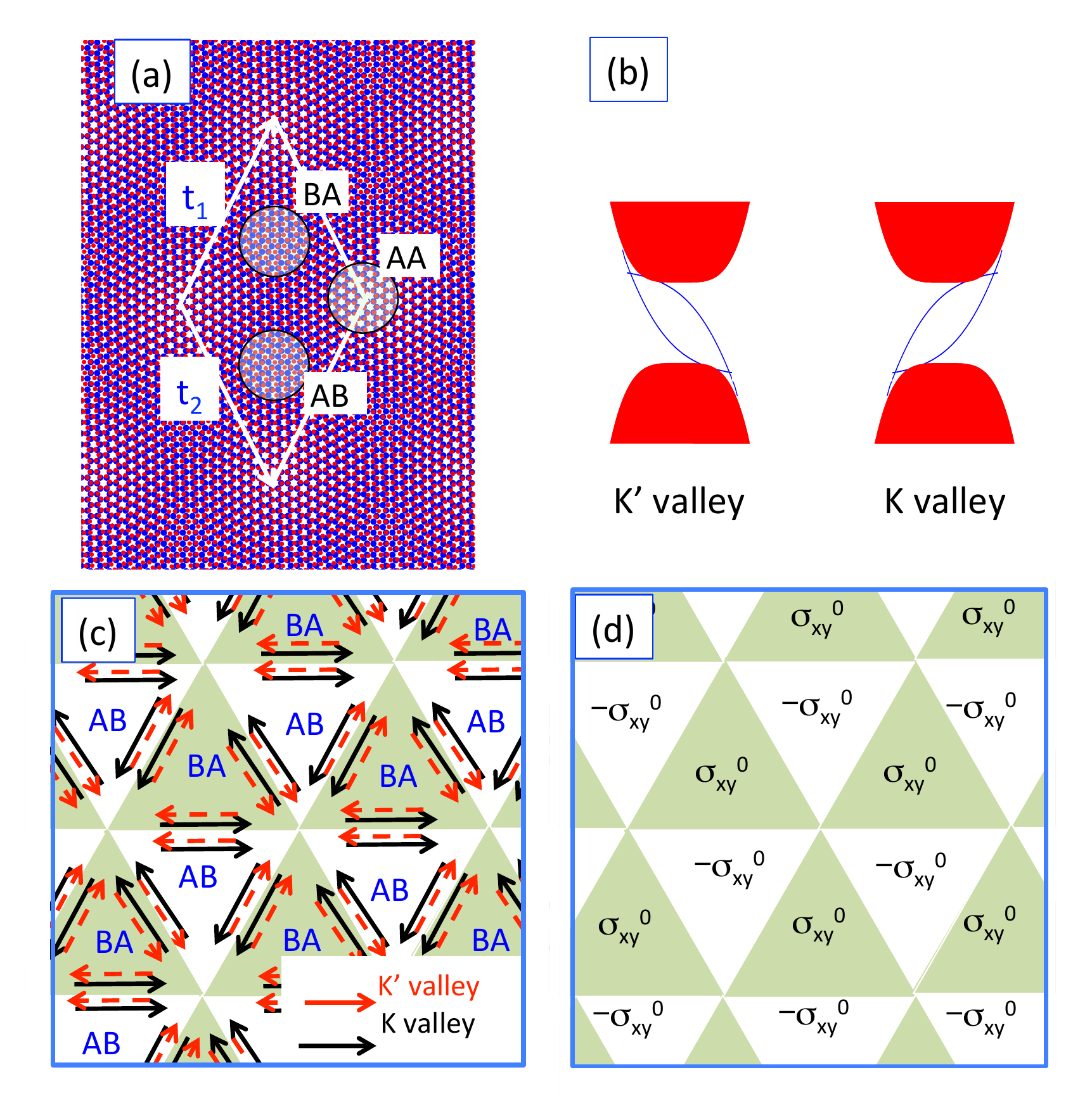}
\caption{(a) TBG's unit cell  showing 
the regions with stacking $AA$, $AB$ and $BA$.
(b) Propagating and anti propagating edge modes corresponding at the $K$ and $K'$ valleys respectively at a  gapped AB/BA interfaces.
(c) Hexagonal pattern showing unit cell with stacking $AB$ and $BA$ and the network of EES encircling these  regions.
(d) Hexagonal pattern  showing the periodicity  of triangular regions with Hall conductivities $\pm \sigma_{xy} ^0$.  }
\label{Figure1}
\end{figure}

The main results of our work are (i) that  plasmons are well represented by CEP bounded to the AB-BA interfaces, that scatter at the hexagon vertices mainly  in the two deflected chiral outgoing directions and (ii)  the prediction of a plasmonic Dirac cone, that is formed by spinors composed of oscillating current patterns in zigzag and armchair direction, respectively. A plasmonic
gap can be opened by breaking the rotational symmetry via strain or  due to
magnetic fields.
Furthermore, the plasmonic density of states diverges in the limit of vanishing frequency. We propose that these features can be seen in scanning near-field microscopy.

\begin{figure}[t]
\includegraphics[width=8.cm]{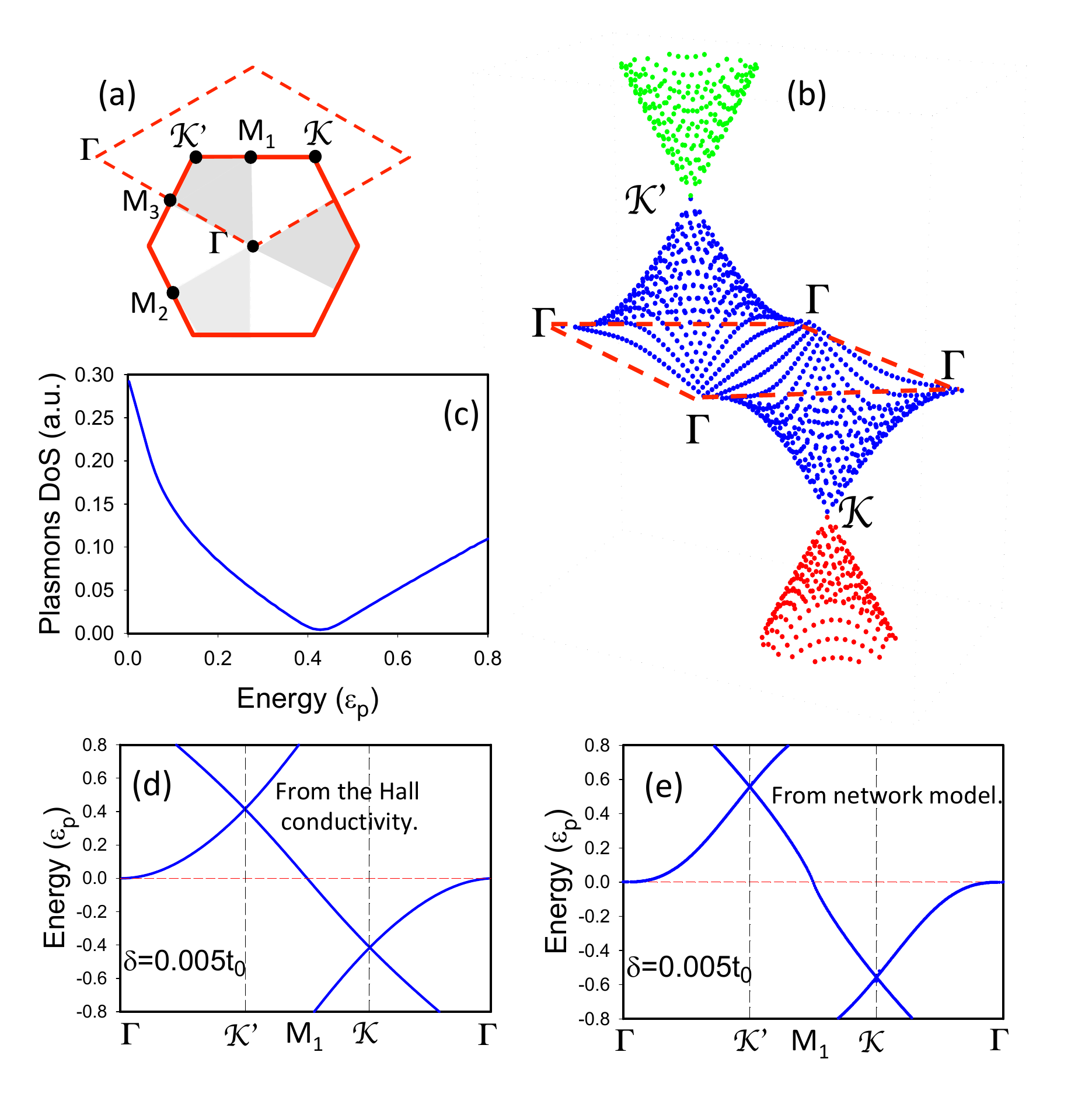}
\caption{(a) 
Moir\'e Brillouin zone. 
Plasmons exit in the shadow regions.
(b) Dispersion of the dielectric modes in the MBZ. Dirac cones appearing at the points ${\cal K}$ and ${\cal K}'$. Red dashed line shows the plane of zero energy. 
(c) Plasmonic density of states.
The flat dispersion in the $\Gamma \! -\! M \! - \! \Gamma$  direction produce a high density of states at low energies. The DoS decays to zero at the frequency of the vertex of the plasmonic Dirac cone. 
(d) Dispersion of the dielectric modes along the direction $\Gamma \! - \! \cal{K}' \! - \! M \! - \! \cal{K} \! - \!  $$\Gamma$ of the MBZ.
(e) As in (d) but obtained with a network model of CEP that scatter at vertices.}
\label{Figure2}
\end{figure} 
\vspace{0.05cm}
\noindent {\it Model.}
The TBG forms a moir\'e pattern that, although in general is incommensurate, can be approximated very accurately by a triangular unit cell of lattice parameters ${\bf t} _{1(2)}$=$\left ( \frac 1 2, \pm \frac {\sqrt{3}} 2 \right ) t_0$, 
with $t_0$=$\frac a {2 	\sin {\theta /2}}$, being $a$=2.46\AA$ $  the graphene lattice parameters and $\theta$ the twist angle, see Fig.\ref{Figure1}(a).  In the TBG unit cell, it is possible to  identify regions where the graphene layers are either Bernal-stacked (AB and BA) or one on top of the other (AA), see Fig.\ref{Figure1}(a).
An interlayer bias, $\Delta$, opens an electronic gap in the Bernal stacked regions.\cite{McCann_2006b} Under lattice relaxation, these regions expand at the expense of the AA regions in order to reduce elastic energy.\cite{Nam:2017aa} For small twist angles, the Bernal-stacked regions define a periodic hexagonal structure, with a unit cell formed by two triangles with stacking BA and AB, respectively, see Fig.\ref{Figure1}(c).

Gapped Bernal-stacked regions are non-trivial insulators and, consequently,  have a finite valley Chern number per spin\cite{Martin_2008,Jung:2011aa,Zhang:2011aa,Prada_2011}, i.e.,
$C_K^{AB}$=$-C_{K'}^{AB}$=$-C_K^{BA}$=$C_{K'} ^{BA}$=${\rm sgn }(\Delta) $, where $K$ and $K'$ are the graphene Dirac points.
At the interface between AB and BA regions, and 
because of the difference in band topology,
two EES per spin and valley appear,\cite{Vaezi:2013aa,Zhang:2013aa,Jaskolski:2018aa} with opposite propagation direction in opposite valleys, which has been experimentally proven.\cite{Ju:2015aa,Yin:2016aa,Li:2016aa,Bor-Yuan17} 
In TBGs the existence of a network of EES encircling the triangular  regions with AB and BA stacking has been predicted\cite{Kindermann12,San-Jose:2013aa,Efimkin:2018aa,hou2019metallic,tsim2020perfect,DeBeule20a,DeBeule20b} and recently experimentally observed.\cite{PhysRevLett.121.037702} 

When the Fermi edge is inside the gapped regions,
the plasmonic 
collective excitations   are formed by electron-hole
electronic transitions  in the  EES and  have an one-dimensional chiral character\cite{Hasdeo:2017aa}.  We name them  chiral edge plasmons (CEP).
In the  context of the quantum Hall effect\cite{Mikhailov:1992aa,Volkov,Shikin,Sommerfeld:1995aa}, CEP exist at the frontier of  two-dimensional  electron gases with Hall conductivities  $ C_1 \frac {e ^2} h $ and $C_2 \frac {e ^2} h$, and have a dispersion of the form $\hbar 
\omega _{1d} ({\bf q})
\approx \frac {e^2 }{2 \epsilon \epsilon_0} q \log (q \delta ) (C_1 \! -\! C_2)$, being $\delta$ the effective thickness of the two-dimensional system. 
A similar dispersion has been obtained for CEP's  at the interface between topological insulators with different Chern numbers\cite{Song:2016aa,Jin:2016aa}. 
In TBG, we expect that because of the Hall conductivities patches, Fig.\ref{Figure1}(d), the collective excitations 
will consist of a network of CEP.

\vspace{0.05 cm}
\noindent {\it Formalism.}
In order to discuss the emerging plasmonic modes, we use the approach based on the current rather than the charge response.\cite{Mikhailov:1992aa} The information on the system is  given by the
conductivity tensor, which in our case has only non-diagonal Hall components with values $\pm \sigma _{xy} ^0=g_s \frac {e^2} h$ in the upper/lower triangles of the moir\'e superlattice, see Fig.\ref{Figure1}(d). Here  $g_s$=2 is   the spin degeneracy. For a wavevector $\bf q$  in the moir\'e  Brillouin zone (MBZ), the  frequencies of the dielectric modes\cite{ Baldereschi-1978} are obtained from the eigenvalues of the matrix
\begin{widetext}
\begin{equation}
M_{{\bf G},{\bf G}'} ({\bf q}) \! = \! - \frac{i}{S_{uc}} \frac {1}{2\epsilon_0\epsilon} \frac { e^{-\frac {\delta} 2  (|{\bf q}+{\bf G}| + |{\bf q}+{\bf G}'|)}}
{\sqrt {|{\bf q}+{\bf G}| |{\bf q}+{\bf G}'| }}
\left [ \hat z \cdot ({\bf q}+{\bf G}) \! \times \! ({\bf q}+{\bf G}') \right ] \tilde {\sigma} _{xy} ({\bf G}-{\bf G}')\;,
\label{matrix}
\end{equation}
\end{widetext}
where $S_{uc}$ and   ${\bf G}$  are the  unit cell area and the reciprocal vectors of the moir\'e lattice, respectively, 
$\tilde {\sigma }_{xy}$ is the Fourier transform of the Hall conductivity, and ${e^{-q \delta}}/({2\epsilon_0\epsilon q})$ is the Fourier transform of the Coulomb interaction in two-dimensions. For a given eigenvalue $\omega _i  ( {\bf q} ) $,  with eigenvector $ \{ \alpha _{\bf q} ^i ({\bf G}) \} $,
the corresponding dielectric eigenmode has the form,
$\phi ^i ( {\bf q}, {\bf G}) = \frac {e^{-\frac {\delta} 2  |{\bf q}+{\bf G}|}} {\sqrt {|{\bf q}+{\bf G}|}} \alpha _{\bf q} ^i ({\bf G})$. 
We have checked that this formalism gives the correct chiral edgeplasmons
when the Hall conductivity is modulated only in one-direction. For details, see the Supplemental Material.\cite{SI}

The matrix $M_{{\bf G},{\bf G}'} ({\bf q})$ is real and symmetric, and therefore its eigenvalues and eigenvectors are real.
The  spectrum is anti-symmetric with respect to zero frequency and zero momentum, i.e., for each eigenvalue $\omega(\bf q)$, there exists a corresponding eigenvalue $\omega (-{\bf q})=-\omega ({\bf q})$ with the same eigenvector. This symmetry guarantees the existence of complex conjugate pairs and the possibility of creating real valued time-dependent electric fields.\cite{Jin:2016aa} Due to the underlying lattice symmetry, we also have
$\omega ({\bf q})=\omega(-q_x,q_y)$=$-\omega(q_x,-q_y)$=$\omega (R_{\frac {2 \pi} 3} {\bf q})$=$-\omega (R_{\frac { \pi} 3} {\bf q})$. 

\vspace{0.05cm}
\noindent {\it Plasmon dispersion.}
Collective charge density excitations of  gapped twisted bilayer graphene
are given by the positive eigenvalues of the matrix $M_{{\bf G},{\bf G}'} ({\bf q})$. 
Because of the chiral nature of the excitations leading to $\omega ({\bf q})=-\omega(-{\bf q})$, they only exist in half of the MBZ, see shadow regions in Fig. \ref{Figure2}(a). The complementary white regions are occupied by plasmonic excitations belonging to the opposite graphene valley. At low energies, the collective excitations are CEP moving along the AB-BA boundaries with a chiral sense of rotation imposed by the arrangement of the Hall conductivity. The  character of the excitations becomes clear when we plot the electrical current associated with a dielectric mode, 
$\vec j ^i ({\bf r})= -\sigma _{xy}({\bf r}) \vec {\nabla} \phi ^i  ({\bf q},{\bf r})$.
In Fig. \ref{Figure3} (a) and (b), we plot the $x$ and $y$ components, respectively, of the electrical current for a mode near $\Gamma$. The current is localized at the sides of the triangles that form the unit cell and  circulates in opposite directions in triangles with opposite Hall conductivity.

\begin{figure}
\includegraphics[width=7.5cm]{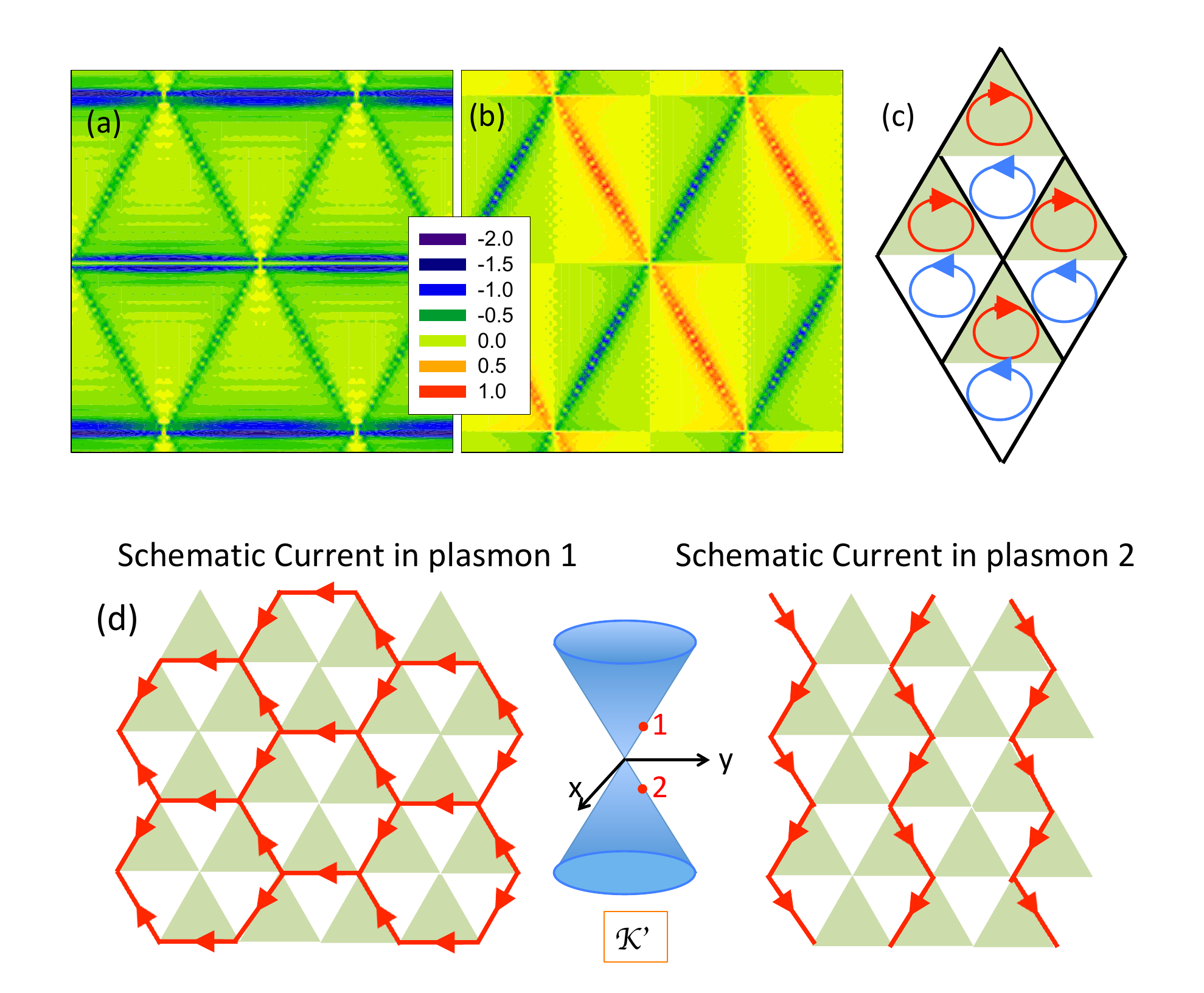}
\caption{Currents in the (a) $x$   and (b) $y$ directions for the low energy plasmons near the $\Gamma$ point of the  MBZ.
(c) Schematic picture of the edge currents corresponding to this plasmon.
Schematic picture of the currents corresponding to the two  plasmons at  the Dirac point ${\cal K}'$. 
The plasmons bring electrical currents in perpendicular directions.}
\label{Figure3}
\end{figure}

In Fig. \ref{Figure2}(d), we plot the first non-zero eigenvalues of  $M_{{\bf G},{\bf G}'} ({\bf q})$ along the direction $\Gamma-{\cal K}'-M-{\cal K}-\Gamma$ of the MBZ. \footnote{In the calculation, we typically use a cutoff for the reciprocal lattice vector with $G_{max}=40\frac{2\pi}{t_0}$ that corresponds to a matrix $M_{{\bf G},{\bf G}'}$ of dimension $4921\times4921$. We have checked the convergence as function of the cutoff.} The most striking feature of the network of CEP dispersion is the existence of a Dirac point at the $\cal K'$ point of the MBZ.
 The dielectric modes for momentum ${\bf q}$, near the $\cal K '$ point, $\bf q$=$\cal K ' +{\bf k}$, can be written in terms of spinors 
$\psi_+=\left ( \begin{array} c \sin {\theta_{\bf k} /2} \\ \cos{ \theta _{\bf k}/2}  \end{array} \right ) e ^{i \bf k  \bf r}$
and $\psi_-=\left ( \begin{array} c \cos {\theta_{\bf k} /2} \\ -\sin{ \theta _{\bf k}/2}  \end{array} \right )   e ^{i \bf k  \bf r} $ with $\theta _{\bf k}$=${ \tan}^{-1}\frac {k_y}{k_x}$. The up and down spinors  are the dielectric modes of the degenerated CEP  at the point ${\cal K '}$. 
Therefore, near  ${\cal K }'$ the CEP are described by the  rotated Dirac equation $H=\hbar v_D(\sigma_zk_x+\sigma_xk_y)+E_D$,
being $v_D$ and $E_D$ the velocity of the plasmons near ${\cal K}'$ and the energy of the plasmons at the vertex of the cone, respectively, see Fig.\ref{Figure2}(c). 
Fig. \ref{Figure3}(d), shows schematically  the electrical currents of the CEP  at the ${\cal K} '$ point, they correspond to  currents moving  along the sides of the triangular networks in the  $\hat x$ (armchair) and $\hat y$ (zigzag) directions. The actual form of the current, as obtained numerically, is plotted in the SM.\cite{SI}

In Fig. \ref{Figure2}(c), we plot the plasmons density of states (DoS)  as function of frequency.
The linear dispersion near the Dirac point
${\cal K}'$ leads to a vanishing DoS at the energy  of the
Dirac cone vertex $E_D$. On the other hand, the $\omega (-{\bf q})=-\omega ({\bf q})$ symmetry, combined with  the chiral
character of the plasmons,  produces zero  energy excitations  along the zone boundaries  of half the MBZ, see Fig. \ref{Figure2}(a)-(b) and a peak in the plasmon DoS at zero frequency, see Fig.\ref{Figure2}(c).

\vspace{0.05cm}

\noindent {\it Plasmons in TBG as a network of CEP.} The currents depicted in  Fig.\ref{Figure3} suggest 
that plasmons in TBG may be described by a network model\cite{Chalker_1988}, similar to that used for
that used to describe the low-energy dispersion relation of the electronic system\cite{Efimkin:2018aa,Pal:2019aa}.  The plasmonic network is composed by the CEP of the AB-BA
 edges  and vertices, where the incoming edge plasmon is scattered in the three possible outgoing directions, 
 see Fig.\ref{Figure1}(c).  This model incorporates Coulomb interactions through the CEP dispersion relation, 
 $\omega _{1d} ({\bf q})$, but neglects the inter-edge interactions. Symmetry and current conservation
 impose that the scattering by a vertex is defined by just two parameters:  (i) the 
 ratio between
 the transmitted current flowing in the forward, $P_{f}$, and each deflected direction,
$P_{d}$, and (ii) the phase of the outgoing forward plasmon, $\phi$.  Remarkably, our
 results\cite{SI} show that the antisymmetry of the plasmon spectrum alone fixes the values $P_{f}
 =1/9$ and $P_{d}=4/9$, and forces $\phi$ to be either 0 or $\pi$.  The parabolicity of the spectrum at the 
 $\Gamma$-point fixes $\phi=\pi$. See Figs.\ref{Figure2}(d) and (e) for a comparison between the plasmon
 dispersion and the one provided by the network model.

\vspace{0.05cm}
\noindent {\it Symmetry breaking.} The existence of a Dirac point indicates the possibility of opening a gap in the CEP  spectrum by, e.g., privileging clockwise electric currents of the
form $J_x+i J_y$ over anti-clockwise currents $J_x-iJ_y$. This can be achieved by applying a magnetic field perpendicularly to the layers.  In that case, an energy gap will appear near $E_D$ and in presence of a magnetic field, biased TBG  will thus become a photonic crystal for nano-light. Also elastic strain should open up a gap by breaking the rotational symmetry. Arguably, the most interesting prospect would be the possibility of creating a Haldane gap leading to topologically protected plasmonic currents, see SM.\cite{SI}


\vspace{0.1cm}
\noindent {\it Energy scale and sound velocity.}
The CEP dispersion is given by only one energy scale, $\epsilon_p=\frac {e^2} {\epsilon_0\epsilon t_0}=\frac{18.1\rm{[eVnm]}}{\epsilon t_0}$, that is determined by the size of the TBF lattice parameter $t_0$. For typical samples with twist angle $\theta\sim0.01^\circ-0.75^\circ$, the corresponding lattice parameter is $t_0\sim19$nm$-1.4\mu$m and the corresponding plasmonic energy scale $\epsilon_p\sim 2.5-200$meV for a dielectric constant $\epsilon \approx 5$, much smaller than the bias voltage/gap of 400meV.\cite{PhysRevLett.121.037702} We thus expect a large energy window where the CEP are not damped by electron-hole excitations, especially not around the Dirac energy $E_D\sim\epsilon_p/2$. 

The plasmonic sound velocity $v_D$ is independent of the moir\'e lattice and can be approximated by $v_D\approx3\sqrt{3}\alpha c/\epsilon$, where $c$ is the speed of light and $\alpha\approx1/137$ the fine-structure constant. 
In general, $v_D << c$, thus justifying our non-retarded electrostatic approach. For
$\epsilon=5$,  $v_D\sim0.0075c$ which is of the same order as the Fermi velocity of graphene.

\vspace{0.05cm}
\noindent
{\it Real space images of the plasmons.} A picture of the electric fields associated with the CEP can be obtained
by using a scanning near field optical microscopy (SNOM) setup\cite{Chen:2012ab,Fei:2012aa,Fei:2013aa}. 
This technique consists in illuminating with an infrared laser the metallic tip of  an atomic force microscope (AFM) placed on top of the TBG. The light induces an electric field dipole at the tip that oscillates with the frequency $\omega$ of the laser and
this oscillation produces an electric field on the underneath TBG.  In order to screen this field, the carriers of the TBG reorganize and induce collective charge excitations, that in the case of gapped TBG are CEP.  These plasmons create an electric field that is backscattered in the tip 
and analyzing the relative variation of the scattering amplitude  as function of the position of the AFM tip it is possible to obtain real-space images of the plasmonics fields.

\begin{figure}[h]
\includegraphics[width=7.5cm]{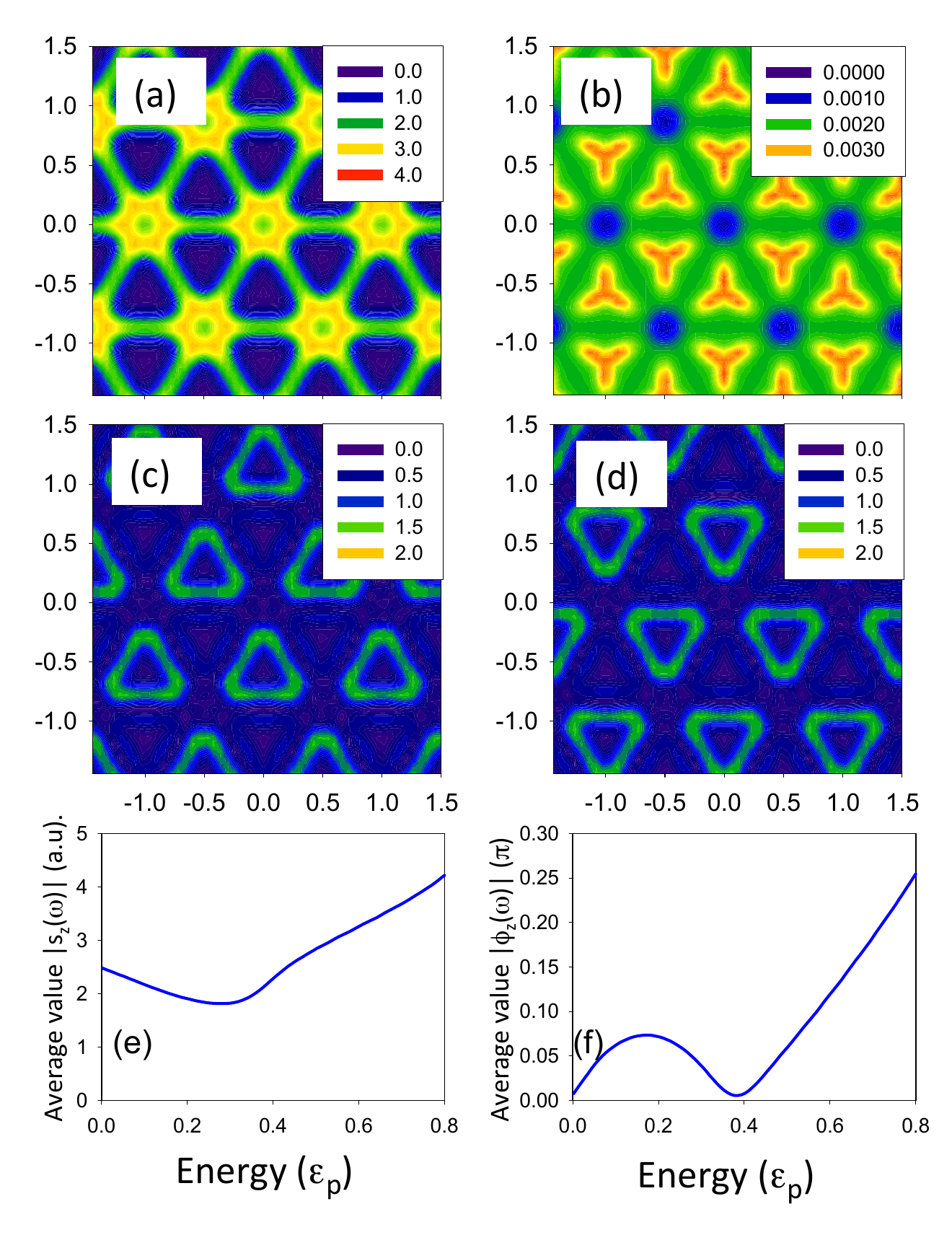}
\caption{Maps of the amplitude of the $z$-component (a) and phase shift (b) of the electric near field.
Maps of the amplitude of the negative (c) and positive (d) circular polarized electric fields.
Variation of the average amplitude (e) and phase shift (f) of the $z$-component of the near electric field.
as function of the frequency. Parameters were chosen to be $z_0$=40nm and  $\delta=0.005t_0$.}
\label{Figure4}
\end{figure}

The tip under illumination can be modeled as an effective point dipole and the electric fields can be evaluated at ${\bf r}\!=\!{\bf r}_0$, see SM.\cite{SI} We obtain real-space images of the electric near-field backscattered by the TBG's plasmons.
In Fig \ref{Figure4} (a)-(b), we plot the map of the amplitudes of the field in the $z$-direction, $s_z (\omega)$, and the corresponding phase shifts $\phi_z(\omega)$ for a frequency $\hbar \omega$=$\epsilon_p/2$. 
Both quantities have six-fold symmetry and they are not sensitive to the chiral nature of the edge states.  The brightest regions of the amplitude of $s_z$ correspond to the corners where the topological edges intersect.  However, the phase shift  of the field in the $z$-direction  is strongest in the middle of the triangles of the unit cell. 
Similar curves are obtained for different frequencies, the main  differences appear in the  values of the amplitude average, $\bar s_z$, and phase shift average, ${\bar {\phi}}_z$ over the unit cell.

In Fig. \ref{Figure4}(e)-(f), 
we plot these quantities as function of the frequency.
${\bar s}_z$ shows a maximum at zero frequency and a minimum at energies slightly  lower than the energy of the
Dirac vertex $E_D$. However, its mathematical expression  is a complicated function of plasmon frequency 
and little knowledge  on the plasmon dispersion can be obtained.
Much more information is obtained from  the average value of the phase shift,
${\bar {\phi}}_z$ which is proportional to the product of the frequency and the plasmon density of states. 
Therefore the zero phase shift 
at the energy $E_D$ indicates the absence of plasmons at this frequency and  shows   the existence of a Dirac cone in the plasmon dispersion.

In Fig. \ref{Figure4}(c) and (d), we plot the real-space images of the amplitudes of the left, $s_-$,  and right, $s_ +$, circular polarized backscattered near electric fields.
Because of the chirality of the  edge plasmons, the electric fields generated by the plasmons  have different chirality on the two triangles of the unit cell.
The electric fields are located at the sides of the triangles of the unit cell.
Upper triangles generate light with negative circular polarization whereas lower triangles produce positive circular polarized light.

\noindent
{\it Summary.}  
We have shown that the collective excitations of biased TBG when the Fermi level lies inside the gap are
chiral edge plasmons that are confined at the AB-BA interfaces.
Contrary to the Drude-like plasmons of Ref. \cite{Sunku:2018aa},  CEP are entirely made  of intra-edge excitations of the 
topological electronic states that occur at the AB-BA edges. Because of their  chiral nature, they only exist in half of the MBZ for one valley and show a large DoS at low frequencies.  
A simple network model shows that scattering of edge plasmons at vertices occurs maximally into the deflected directions. Most strikingly, we observe a plasmonic Dirac cone at finite energy that is composed of oscillating currents in $x$ and $y$-direction. Breaking the rotational symmetry by a magnetic field opens up a tunable gap, paving  the way towards a plasmonic transistor at tuneable frequencies $E_D\sim1 \! - \! 100$meV depending on the twist angle. 

\noindent
{\it Acknowledgments}. This work has been supported by Spain's MINECO under Grant No. PGC2018-097018-B-100, PGC2018-096955-B-C42, FIS2017-82260-P, and MAT2017-88358-C3-1-R as well as by the CSIC Research Platform on Quantum Technologies PTI-001 and the European Union Seventh Framework Programme under grant agreement no.785219 and no. 881603 Graphene Flagship for Core2 and Core3.  LMM acknowledges Arag\'on Government through project Q-MAD.

\pagebreak
 
\begin{widetext}
\vspace{1.cm}

\centerline {\large {SUPPLEMENTAL MATERIAL Plasmonic Dirac Cone  in Twisted Bilayer Graphene }}
\vspace{0.3cm}
\centerline {\large {L.Brey, T.Stauber, T.Slipchenko and L. Mart\'{\i}n-Moreno }}

\section{Collective excitations of a chiral Hall network}
Here, we outline the formalism how to obtain the collective excitations of a two-dimensional system characterized by a spatial modulation of the Hall conductivity  and zero longitudinal conductivity. The input in  the calculation is  the electrical conductivity tensor,
\begin{equation}
\sigma ({\bf  r})= \left ( \begin{array}{cc} \sigma _{xx} & \sigma_{xy}({\bf r}) \\
\sigma_{yx} ({\bf r}) &\sigma_{yy} \end{array} \right )
\end{equation}
with $\sigma_{xx}$=$\sigma_{yy}$=$0$, and
\begin{equation}
\sigma_{xy}({\bf r})=-\sigma_{yx}({\bf r})=\frac 1 S \sum _{\bf G} \tilde \sigma_{xy}({\bf G}) e ^{i {\bf G}{\bf r}}\;.
\end{equation}
Here, $S$ is the unit cell area and the sum runs over all reciprocal lattice vectors associated with the periodicity of the conductivity in real space.  We consider an external potential acting on the system,
\begin{equation}
\phi _{ext} ({\bf  r} )= \tilde \phi _{ext} ({\bf q})  e ^{i {\bf q} {\bf r}} e ^{i \omega t} \, ,
\end{equation}
here the wavevector is restricted to the Brillouin zone of the periodic structure. 
Due to the periodicity of the conductivity, the total potential in the system has the general form,
\begin{equation}
\phi _T ({\bf r}) = e ^{i {\bf q} {\bf r}} e ^{i \omega t} \sum _{\bf G} \tilde \phi _T({\bf G},{\bf q})e ^{i {\bf G}{\bf r}}\;.
\end{equation}
The electric field associated with this potential,
\begin{equation}
{\bf E}({\bf r})=- {\pmb {\nabla}} \phi _T ({\bf r}) 
\end{equation}
induces, we assume locally, an electrical current density
\begin{equation}
{\bf J}({\bf r})= \sigma ({\bf r}) {\bf E} ({\bf r})
\end{equation}
that is related with the two dimensional electron density trough the continuity equation,
\begin{equation}
\frac {\partial \rho ({\bf r})} {\partial t} + {\pmb {\nabla}} \cdot {\bf J} ({\bf r})=0 \, \, .
\end{equation}
Finally the induced charge density creates a Hartree potential that in the spirit of the RPA should be added to the external potential in order to 
obtain the following  selfconsistency equation,
\begin{equation}
\tilde \phi _T ({\bf G},{\tilde q})= \tilde \phi _{ext} ({\tilde q})+ v({\bf q} +{\bf G}) \frac 1 {i \omega} \frac 1 S  
\sum _{\bf G'} {\bf \hat z} \cdot ({\bf q}+{\bf G})\times  ({\bf q}+{\bf G}') \tilde \sigma _{xy} ({\bf G}-{\bf G}')\, \tilde \phi _T ({\bf G}',{\tilde q})
\end{equation}
where $v({\bf q})= \frac {1}{2\epsilon_0\epsilon q} e ^{-\delta q}$ is the two-dimensional Fourier transform of the Coulomb potential and $\delta$ is the effective thickness of the two-dimensional electron gas. 
From this equation we obtain the matrix dielectric constant in reciprocal space,
\begin{equation}
\epsilon _{{\bf G},{\bf G}'} ({\bf q})=
\delta _{{\bf G},{\bf G}'} - \frac 1 S \frac 1 {i \omega} \frac {1}{2\epsilon_0\epsilon |{\bf q}+{\bf G}|} e^{-\delta |{\bf q}+{\bf G}|} \left [ {\bf \hat z} \cdot ({\bf q}+{\bf G})\times  ({\bf q}+{\bf G}') \right ]  \tilde \sigma _{xy} ({\bf G}-{\bf G}')
\end{equation}
Plasmons are self-sustained charge excitation in the system, that occur when the dielectric constant is zero for a particular frequency $\omega$ and
momentum  ${\bf q}$. The zero eigenvalues of the matrix $\epsilon _{{\bf G},{\bf G}'} ({\bf q})$ give us the dispersion $\omega ({\bf q})$ of the plasmon in the system.
The dielectric mode  associated with a particular plasmon $\omega ({\bf q})$, is given by the eigenvector of the matrix 
$\epsilon _{{\bf G},{\bf G}'} ({\bf q})$ corresponding to the zero eigenvalue.

It is more convenient for numerical computation  to define a symmetrized matrix
\begin{equation}
\epsilon  ^{\prime} _{{\bf G},{\bf G}'} ({\bf q})=   \sqrt{ |{\bf q}+{\bf G}|} \, \epsilon _{{\bf G},{\bf G}'} ({\bf q})  \frac 1 {\sqrt{ |{\bf q}+{\bf G'}|}}  
e^{-\frac {\delta} 2(|{\bf q}+{\bf G}|-|{\bf q}+{\bf G}'|)}
\end{equation}
It is easy to prove that the eigenvalues of $\epsilon  ^{\prime} _{{\bf G},{\bf G}'} ({\bf q})$ correspond to the eigenvalues of 
$\epsilon   _{{\bf G},{\bf G}'} ({\bf q})$.
The eigenvectors $\alpha _{{\bf q}} ({\bf G})$ of $\epsilon  ^{\prime}$ are related with the eigenvectors $\phi ({\bf q},{\bf G}) $ of $\epsilon$ through,
\begin{equation}
\alpha _{{\bf q}} ({\bf G}) =  \sqrt{ |{\bf q}+{\bf G}|}  \, e^{\frac {\delta} 2|{\bf q}+{\bf G}|}  \, \phi ({\bf q},{\bf G})
\end{equation}
The Hall conductivity does not depend on frequency, and in the dielectric tensor the frequency only appear as $\frac 1 {\omega}$. therefore the zeros of the dielectric constant  are the eigenvalues of the matrix,
\begin{equation}
M_{{\bf G},{\bf G}'} ({\bf q}) \! = \! - \frac i {S} \frac {1}{2\epsilon_0\epsilon} \frac { e^{-\frac {\delta} 2  (|{\bf q}+{\bf G}| + |{\bf q}+{\bf G}'|)}}
{\sqrt {|{\bf q}+{\bf G}| |{\bf q}+{\bf G}'| }}
\left [ \hat z \cdot ({\bf q}+{\bf G}) \! \times \! ({\bf q}+{\bf G}') \right ] \tilde {\sigma} _{xy} ({\bf G}-{\bf G}')
\label{matrix}
\end{equation}
$\tilde \sigma _{xy} (n {\bf G}_1+ m{\bf G}_2) $ is  pure imaginary so that the matrix 
$M_{{\bf G},{\bf G}'} ({\bf q})$ is real and symmetric and have real eigenvalues and eigenfunctions. Note that the plasmonic energy $\hbar\omega$ only depends on the energy scale $\epsilon_p=\frac{e^2}{\epsilon_0\epsilon  t_0}$.

\section{Fourier Transform of $\sigma _{xy}( {\bf r})$ }
The supercell of the twisted bilayers is defined by the lattice vector 
$\bf t _1$=$\left ( \frac 1 2, \frac {\sqrt{3}} 2 \right ) t_0$ and $\bf t _2$=$\left ( \frac 1 2, -\frac {\sqrt{3}} 2 \right ) t_0$ ,with $t_0$=$\frac a {2 	\sin {\theta /2}}$, being $a$=0.246$nm$ the graphene lattice parameters and $\theta$ the twist angle, see Fig.\ref{Figure1}(a) of the main text.
The reciprocal primitive vectors are
\begin{eqnarray}
{\bf G}_1 &=& \frac {4 \pi}{\sqrt {3} t_0}\left ( -\frac {\sqrt{3}} 2, \frac 1 2 \right ) \\
{\bf G}_2 &=& \frac {4 \pi}{\sqrt {3} t_0}\left ( \frac {\sqrt{3}} 2, \frac 1 2 \right ) \, \, .
\end{eqnarray}
For a reciprocal lattice vector ${\bf G}=n {\bf G}_1+ m{\bf G}_2$ the Fourier transform of 
$\sigma _{xy} ({\bf r})$ as shown in Fig.1(d) of the main text, is
\begin{eqnarray}
\tilde \sigma _{xy} (n {\bf G}_1+ m{\bf G}_2) & =& -\sigma _{xy} ^0 \frac {\sqrt{3}}{2 n \pi} i t_0^2 \, \,  {\rm if} \, \,  n=m \ne 0  \nonumber \;,\\ 
\tilde \sigma _{xy} (n {\bf G}_1+ m{\bf G}_2) & =& \sigma _{xy} ^0 \frac{\sqrt{3}}  {2 n \pi} i t_0^2 \,  \,\,\,\, \, {\rm if} \, \, m=0 \, {\rm and} \,  n \ne 0   \nonumber\;, \\ 
\tilde \sigma _{xy} (n {\bf G}_1+ m{\bf G}_2) & =& \sigma _{xy} ^0 \frac{\sqrt{3}} {2 m \pi} i t_0^2 \, \, \,\,\, {\rm if} \, \,n=0 \, {\rm and} \,  m \ne 0   \nonumber\;,
\end{eqnarray}
and $\tilde \sigma _{xy} (n {\bf G}_1+ m{\bf G}_2)=0$ otherwise. Note that with $\sigma_{xy}^0=2\frac{e^2}{h}$, the Fourier transform $\tilde \sigma _{xy}$ is purely imaginary such that $M_{{\bf G},{\bf G}'} ({\bf q})$ of Eq. (\ref{matrix}) is real and symmetric. 

\section{Edge chiral plasmons for one-dimensional modulation.}
\begin{figure}[h]
\includegraphics[width=10cm]{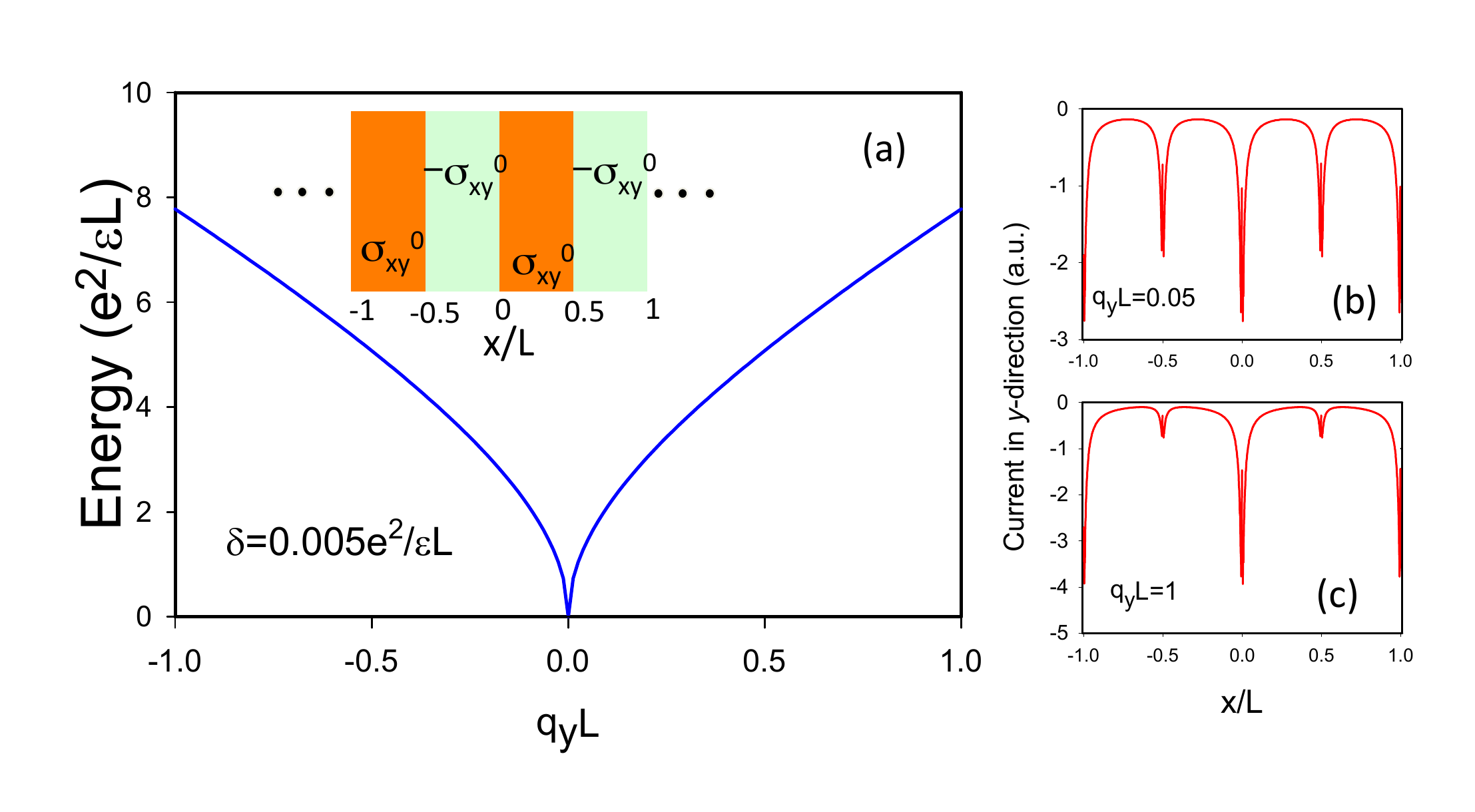}
\caption{(a) Dispersion of the edge plasmons in a system with one-dimensional modulation of  the Hall conductivity. 
In the inset it is shown schematically the modulation of $\sigma _{xy}$. In panels (b) and (c) we plot the electrical currents associated at the 
edge plasmons with momentum $q_y L$=0.05, and $q_y L$=1.
}
\label{Figure1}
\end{figure} 

In this section, we study a two-dimensional system where the Hall conductivity is periodic only in the $\hat x$-direction, with a period $L$, see Fig.\ref{Figure1},
\begin{eqnarray} 
\sigma _{xy} & = & -\sigma_{xy} ^0 \,\,\, {\rm for} \,\,\, x<L/2 \\
\sigma _{xy} & = & \sigma_{xy} ^0 \,\,\, {\rm for} \,\,\, x>L/2 \, \, 
\end{eqnarray}
Due to the change in the Hall conductivity, it is expected that the system supports chiral edge currents moving on
opposite directions at the interfaces $-\sigma _{xy} ^0$/$\sigma _{xy} ^0$
than in the neighbor interfaces  $\sigma _{xy} ^0$/$-\sigma _{xy} ^0$.

The Fourier transform has only components in the $\hat x$direction,
\begin{equation}
\tilde \sigma (n G_0)= -i  \sigma _{xy} ^0 \frac L {\pi n} \left ((-1)^n-1 \right)
\end{equation}
with $G_0=\frac {2 \pi }{L}$.
The matrix $M$ becomes,
\begin{equation}
M _{G_x,G'_x} (q_y,\omega) \! = \! - \frac i {L} \frac {1}{2\epsilon_0\epsilon} \frac { e^ {-\frac {\delta} 2  (\sqrt{q_y ^2+ G_x^2} +\sqrt{q_y^2+G_x'^2}) } }
 {\sqrt { \sqrt{q_y ^2+ G_x^2} \sqrt{q_y^2+G_x'^2} }}
\, q_y  \,  ( G_x-G_x' ) \, \tilde {\sigma} _{xy} (G_x-G_x')
\end{equation}

In Fig.\ref{Figure1} we plot the dispersion relation corresponding to the edge plasmons. At small wavevectors the plasmons 
disperse as
$\sqrt {q}$, this behavior  occurs because for wavevectors much smaller than $1/L$  the plasmons average over all the one-dimensional channels and the system behaves as a two-dimensional system.
In  Fig.\ref{Figure1}(b) we plot the current along the edges direction, the current is parallel in both interfaces of the unit cell,  this 
is a reflection of the 
Coulomb coupling between edges at small wavevectors. The
excitations can still be classified as chiral 
but now a mode propagating on an edge is dragged by the
charge on the other edge\cite{Franco:1996aa}. At larger values of the momentum the coupling between edges is reduced and  the current is practically carried by   the  channels  at the interfaces   $-\sigma _{xy} ^0$/$\sigma _{xy} ^0$, see Fig.\ref{Figure1}(c).

At large wavevectors the dispersion relation of Fig.\ref{Figure1}(a) can be described as the coupling between  edge modes with the same chirality,
\begin{equation}
\hbar \omega = \frac {n_c}{2 \pi} q_x \frac {e^2}{2 \epsilon_0\epsilon} \left  ( K_0 (|q_x| a)+ \sum _{n=2} ^{\infty} K_0 (|q_x| n L) \, \right ) \;.
\end{equation}

\section{Chiral Network model}

\begin{figure}[t]
\includegraphics[width=12cm]{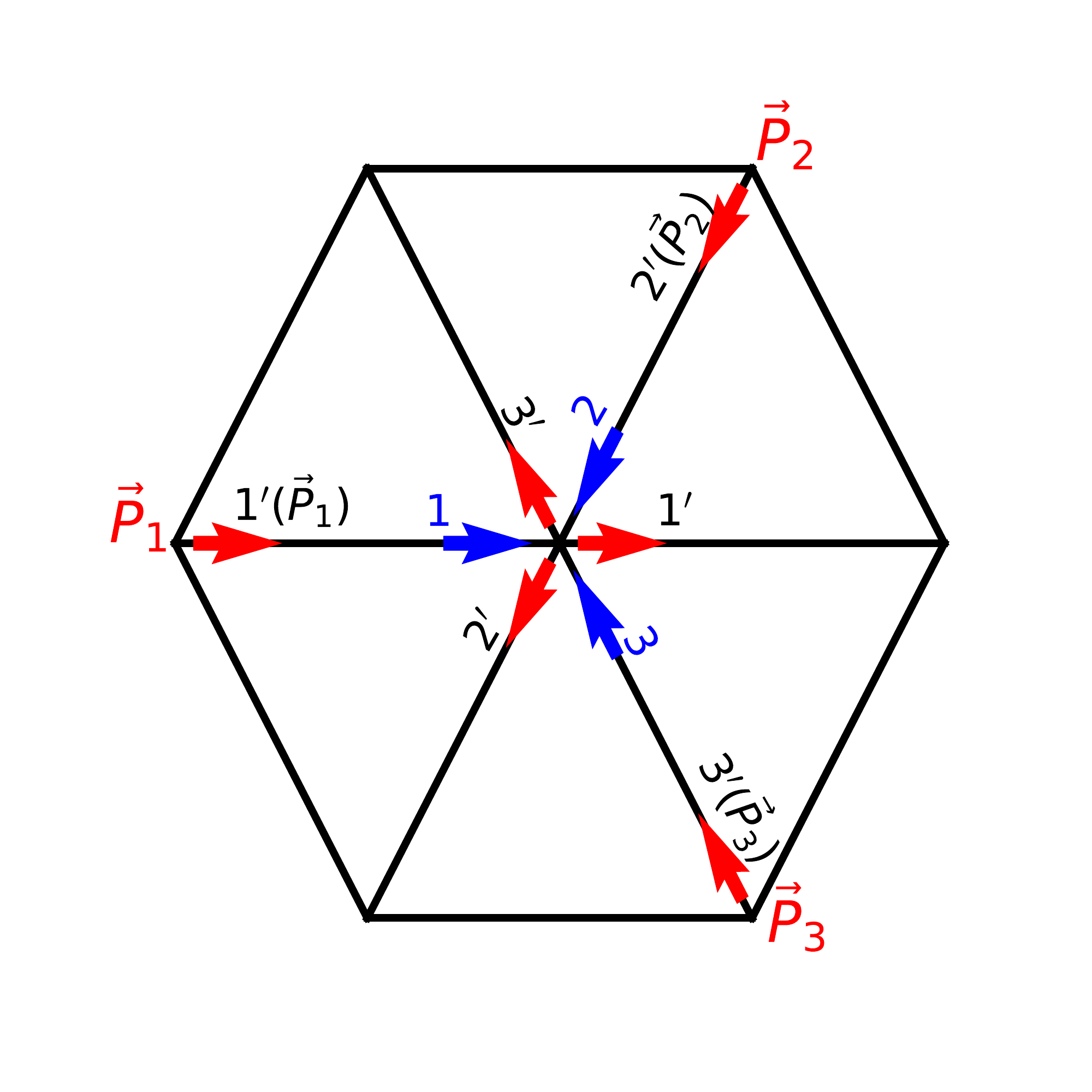}
\caption{Relevant directions in the chiral network model. Red arrows represent outgoing fields from a vertex at lattice position $\vec{R}$, with the three chiral outgoing directions labelled as $1^{\prime}, 2^{\prime}$ and $3^{\prime}$ (for simplicity the label corresponding to $\vec{R}=0$ has been omitted). Blue arrows represent incoming fields at a vertex.}
\label{vertex}
\end{figure} 

\begin{figure}[h!]
\includegraphics[width=12cm]{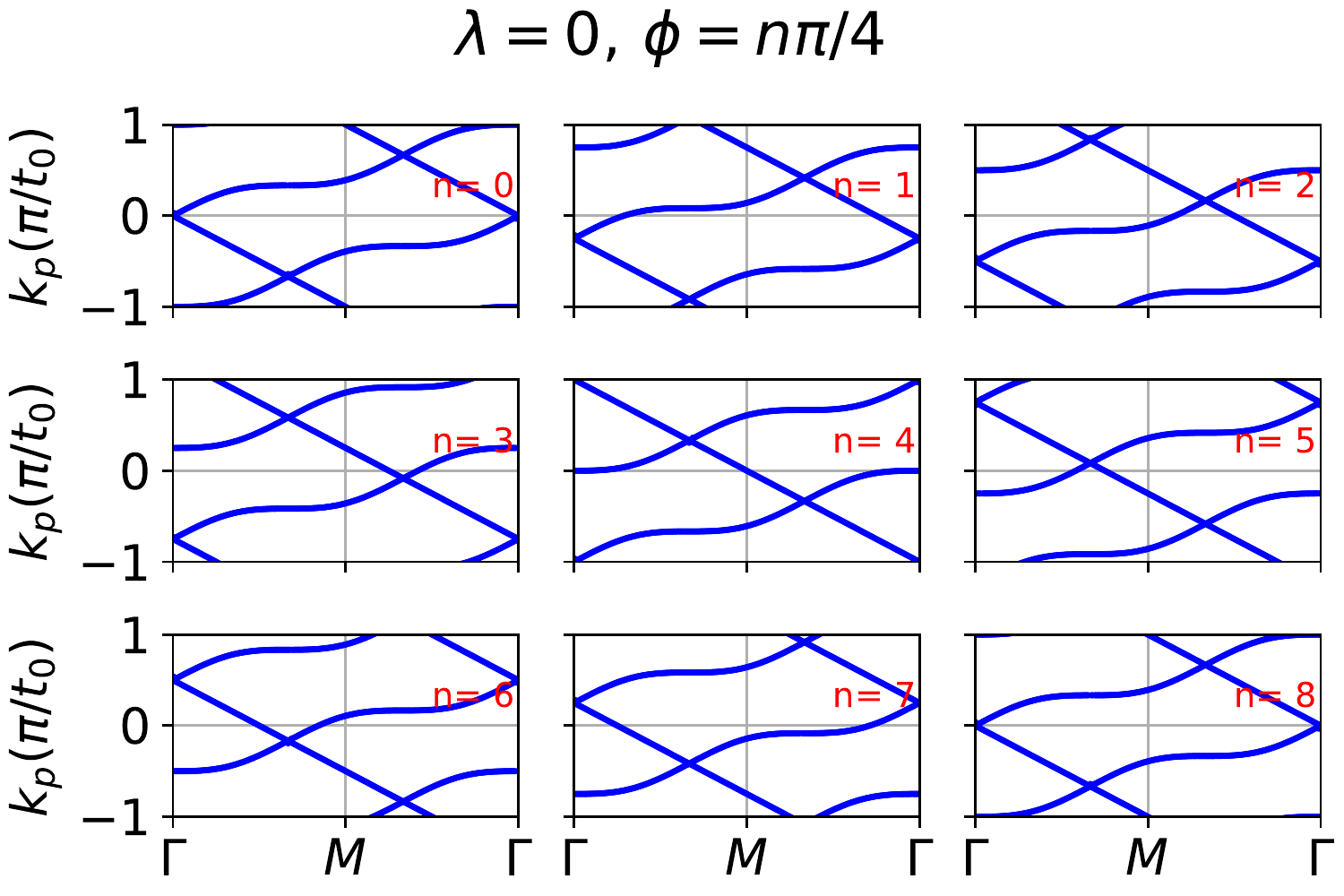}
\caption{Dispersion relation in a network model of chiral 1D-plasmons for several values of $\phi$, at a fixed representative value of $\lambda=0$. The results are periodic on $\phi$ with period $2\pi$, as shown by the panels $n=0$ and $n=8$.}
\label{landa0}
\end{figure} 

\begin{figure}[h!]
\includegraphics[width=12cm]{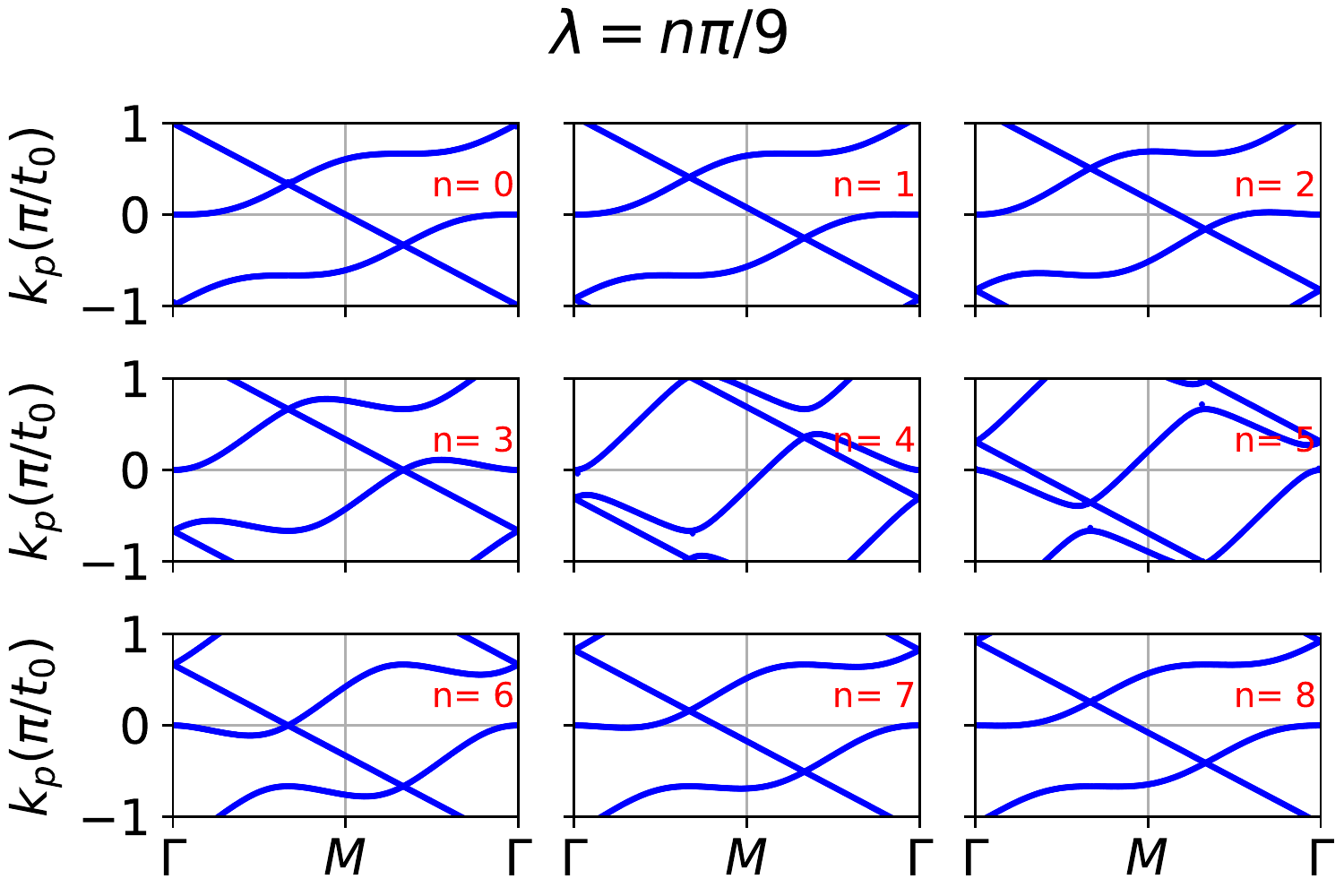}
\caption{Dispersion relation in a network model of chiral 1D-plasmons,  for several values of $\lambda$. In each panel the phase $\phi$ has been chose so that $k_{p}(\Gamma) = 0$}.
\label{fi0}
\end{figure} 

A simplified model for the plasmons in a TBG is the chiral network model, which was already used to discuss the electronic properties in TBG\cite{Efimkin:2018aa}.  We present here a short summary of the model and adapt it to the plasmonic case. 

The network model is constructed by quasi-one-dimensional chiral waves that propagate along the hexagon lines, and scatter at the hexagon vertices. The propagation along the lines is characterised by the wavevector of the waves at the considered frequency. In the plasmonic case, the chiral waves are the plasmons bounded to the one-dimensional AB-BA edge. Their dispersion relation is 
$\hbar \omega _{1d} ( k_{p}) \approx 4 e^2 k_{p} \log (k_{p} \delta )$, being $k_{p}$ the plasmon wavevector along the propagation direction and $\delta$ the effective thickness of the 2D system (see main text).

The network model assumes that the scattering at the vertex conserves valley index.  Then, for a given valley index, chiral plasmons can approach the vertex from three incoming directions (labelled 1, 2 and 3) and leave the vertex by three outgoing ones (labelled ${1}^{\prime} ,{2}^{\prime}$,  and ${3}^{\prime}$), see Fig.\ref{vertex} for a schematic representation.  Although the plasmon is a vector field,  it is sufficient to consider the out-of plane component of the electric field ($E_{z}$), as all components of the electromagnetic field can be extracted from it using Maxwell equatuons.  At a vertex $\vec{R}$, we denote the amplitude of the outgoing field  by  $\vec{E}^{\prime}_{z}(\vec{R}) \equiv \begin{pmatrix} E_{z}^{1\prime}(\vec{R}), E_{z}^{2\prime}(\vec{R}),E_{z}^{3\prime}(\vec{R}) \end{pmatrix}^{T}$, and the  amplitude of the incoming field by $\vec{E}_{z}(\vec{R}) \equiv \begin{pmatrix} E_{z}^{1}(\vec{R}), E_{z}^{2}(\vec{R}),E_{z}^{3}(\vec{R}) \end{pmatrix}^{T}$, where $^{T}$ stands for transpose, and the superindex $E^{i}_{z}$ indicates the amplitude of $E_{z}$ in the i-th direction. The scattering matrix ${\widetilde {S}}$ relates the outgoing amplitudes to the incoming ones thorough 
\begin{equation}
\vec{E}_{z}^{\prime}(\vec{R}) = {\widetilde {S}} \, \vec{E}_{z}(\vec{R}). 
\end{equation}
The scattering matrix must be both unitary (due to current conservation) and symmetrical with respect to the two deflected outgoing directions. It is determined by just two parameters:  (i) $\phi$, the phase picked up in the scattering process by the plasmon propagating in the outgoing forward direction and (ii) the fraction of power in the forward direction $P_{f}$ (the fraction of power in a deflected direction, $P_{d}$, satisfies $P_{f} + 2P_{d} =1$).  We use the parametrisation for ${\widetilde {S}}$ presented in \cite{Pal:2019aa}:
\begin{equation}
{\widetilde {S}} (\phi, \lambda) = e^{\imath \phi} \begin{pmatrix} \alpha & \beta e^{\imath \lambda} & \beta e^{\imath \lambda} \\
\beta e^{\imath \lambda}&\alpha & \beta e^{\imath \lambda} \\
\beta e^{\imath \lambda}&\beta e^{\imath \lambda}&\alpha 
\end{pmatrix}
\end{equation}
where $\alpha = 1/\sqrt{1+8\cos^{2}\lambda}$, $\beta = -2\cos\lambda/\sqrt{1+8\cos^{2}\lambda}$,  and $P_{f} =\alpha^{2}$, $P_{d} =\beta^{2}$.

Of course, all vertices in the lattice are equivalent; for simplicity let us focus on the vertex placed at $\vec{R}=0$.
The equations governing the plasmons in the network can be found by noticing, first, that the incoming amplitudes at $1, 2$ and $3$ are related to the outgoing amplitudes at $\vec{P}_{1}= - (\vec{P}_{2}+ \vec{P}_{3})$, $\vec{P}_{2}  = (1, -\sqrt{3})^{T} t_{0}/2$ and $\vec{P}_{3}  = (1, \sqrt{3})^{T} t_{0}/2$ (see Fig.\ref{vertex}) through 
\begin{equation}
\begin{pmatrix} E_{z}^{1}(0) \\E_{z}^{2}(0) \\ E_{z}^{3}(0) \end{pmatrix} =
\underbrace{\begin{pmatrix}  \Psi &0&0 \\ 0&\Psi & 0\\ 0&0&\Psi \end{pmatrix}}_{\widetilde{P}(k_{p})}
\begin{pmatrix} E_{z}^{1\prime}(\vec{P}_{1}) \\ E_{z}^{2\prime}(\vec{P}_{2}) \\E_{z}^{3\prime}(\vec{P}_{3}) \end{pmatrix}
\end{equation}
where $\Psi = e^{\imath k_{p} t_{0}}$ is the phase that a plasmon acquires when propagating between two vertices. Second, that according to Bloch's theorem 
\begin{equation}
\begin{pmatrix} E_{z}^{1\prime}(\vec{P}_{1}) \\ E_{z}^{2\prime}(\vec{P}_{2}) \\E_{z}^{3\prime}(\vec{P}_{3}) \end{pmatrix}
 =
\underbrace{\begin{pmatrix}  e^{\imath \vec{k} \vec{P}_{1}} &0&0 \\ 0&e^{\imath \vec{k} \vec{P}_{2}}& 0\\ 0&0&e^{\imath \vec{k} \vec{P}_{3}} \end{pmatrix} }_{\widetilde{B}(\vec{k})}
\begin{pmatrix} E_{z}^{1\prime}(0) \\ E_{z}^{2\prime}(0) \\E_{z}^{3\prime}(0) \end{pmatrix}.
\end{equation}

Thus, the plasmonic modes satisfy the equation 
\begin{equation}
\widetilde{\Omega} \begin{pmatrix} E_{z}^{1\prime}(0) \\ E_{z}^{2\prime}(0) \\E_{z}^{3\prime}(0) \end{pmatrix} = \begin{pmatrix} E_{z}^{1\prime}(0) \\ E_{z}^{2\prime}(0) \\E_{z}^{3\prime}(0) \end{pmatrix}
\end{equation}
where $\widetilde{\Omega}(k_{p}, \vec{k}, \phi, \lambda) \equiv \widetilde{S}(\phi, \lambda) \widetilde{P}(k_{p}) \widetilde{B}(\vec{k}) $. Thus, for given Bloch wavevector $\vec{k}$, this eigenvalue equation can be readily solved by finding the values of $k_{p}$ for which $\widetilde{\Omega}(k_{p}, \vec{k}, \phi, \lambda) $ has an eigenvalue equal to 1.  The dispersion relation of the network $\omega(\vec{k})$ is obtained by using the dispersion relation of the quasi-one-dimensional mode $\omega_{1d}(k_{p}(\vec{k}))$. 

As the relation between $\omega$ and $k_{p}$ practically linear in the range of interest,  we concentrate on $k_{p}(\vec{k})$, which has the mathematical property that if $k_{p}$ is a solution, so is $k_{p} + 2 \pi n$, with $n$ being an integer. So, only the values $\k_{p} \in [-\pi, \pi]$ need to be considered.

For completeness, we show the general dependence of the band structure on the two parameters defining the scattering matrix, $\phi$ and $\lambda$.  
The dependence on $\phi$, for a representative $\lambda=0$,  is shown in Fig.\ref{landa0}, illustrating that the variation of $\phi$ produces a rigid vertical displacement in the $k_{p}(\vec{k})$ dispersion relation. The dependence on $\lambda$ is rendered in Fig.\ref{fi0}, showing that $\lambda$ changes the overall shape of the dispersion relation (for instance, the quasiparticle mass at $\Gamma=0$).

The numerical results could be used to fit the TBG plasmonic bands rendered in Fig.2 of the main text, obtaining in this way the parameters $\phi$ and $\lambda$ that define the chiral plasmon scattering at the vertex.
However, the obtention of these parameters is simplified by
\begin{itemize} 
\item[(i)]
using the antisymmetry of the plasmonic spectrum discussed in the main text, i.e. $\omega(\vec{k}) = - \omega(-\vec{k})$, or $k_{p}(\vec{k}) = - k_{p}(-\vec{k})$ and, 
\item[(ii)]
noticing that 
$\widetilde{B}^{*}(\vec{k}) = \widetilde{B}(-\vec{k})$ and $\widetilde{P}^{*}(k_{p}) = \widetilde{P}(-k_{p})$
\end{itemize}

In order to use these symmetries, notice that any solution satisfies $ \widetilde{\Omega}(k_{p}, \vec{k}, \phi, \lambda) \vec{v} = \vec{v}$, and taking the complex conjugate we get $ \widetilde{\Omega}^{*}(k_{p}, \vec{k}, \phi, \lambda) \vec{v}^{*} = \vec{v}^{*}$. Using the properties of $\widetilde{B}$ and $\widetilde{P}$ this leads to
\begin{equation}
\widetilde{S}^{*}(\phi, \lambda) \, \widetilde{P}(-k_{p}) \, \widetilde{B}(-\vec{k})  \, \vec{v}^{*} = \vec{v}^{*},
\label{eqS}
\end{equation}
so the antisymmetry of $k_{p}(\vec{k})$ is fulfilled if $\widetilde{S}^{*}(\phi, \lambda) =\widetilde{S}(\phi, \lambda) $, when Eq. (\ref{eqS}) becomes $\widetilde{\Omega}(-k_{p}, -\vec{k}, \phi, \lambda) \, \vec{v}^{*} = \vec{v}^{*}$, which also determines an eigenvalue equal to unity and thus the existence of a plasmonic mode.
Therefore, the antisymmetry of the spectrum implies that $\widetilde{S}$ should be real-valued, this is, both $\phi$ and $\lambda$ should be either 0 or $\pi$ (plus an irrelevant multiple of $2 \pi$).  In fact, $\lambda=0$ and $\lambda=\pi$ correspond to the same scattering matrix, which fixes $\alpha =  1/3$,  $\beta = 2/3$, and thus the forward and defected currents $P_{f}=1/9$ and $P_{d}= 4/9$, respectively.  So, symmetry alone forces that the scattered current in the forward direction has its minimum possible value (which, remarkably, is finite). 

The value of $\phi$ can be found by comparing the bands for $\phi=0$ and $\phi=\pi$ in Fig.\ref{landa0} with those in Fig.2 of the main text. Notice that, as expected, both cases give antisymmetry bands but only $\phi=\pi$ leads to a parabolic band at $\Gamma=0$. 

To summarise, the antisymmetry of the bands force $\lambda = 0$, and this plus the parabolic dispersion at $\Gamma$ forces $\phi=\pi$.

\section{Electric field induced by an oscillatory dipole.}
When illuminating the metallic tip of an atomic force microscope with a focused infrared laser beam, it is created
an electric dipole ${\bf P}$  that oscillates with the frequency oscillating with frequency $\omega$  of the laser.
We assume that the radiation of the tip can be simulated by a point dipole situated at  position 
${\bf r }_0$, see Fig.\ref{Figure2}.  This dipole is going to excite the collective modes of bilayer graphene inducing an electric field.

\begin{figure}[h]
\includegraphics[width=8cm]{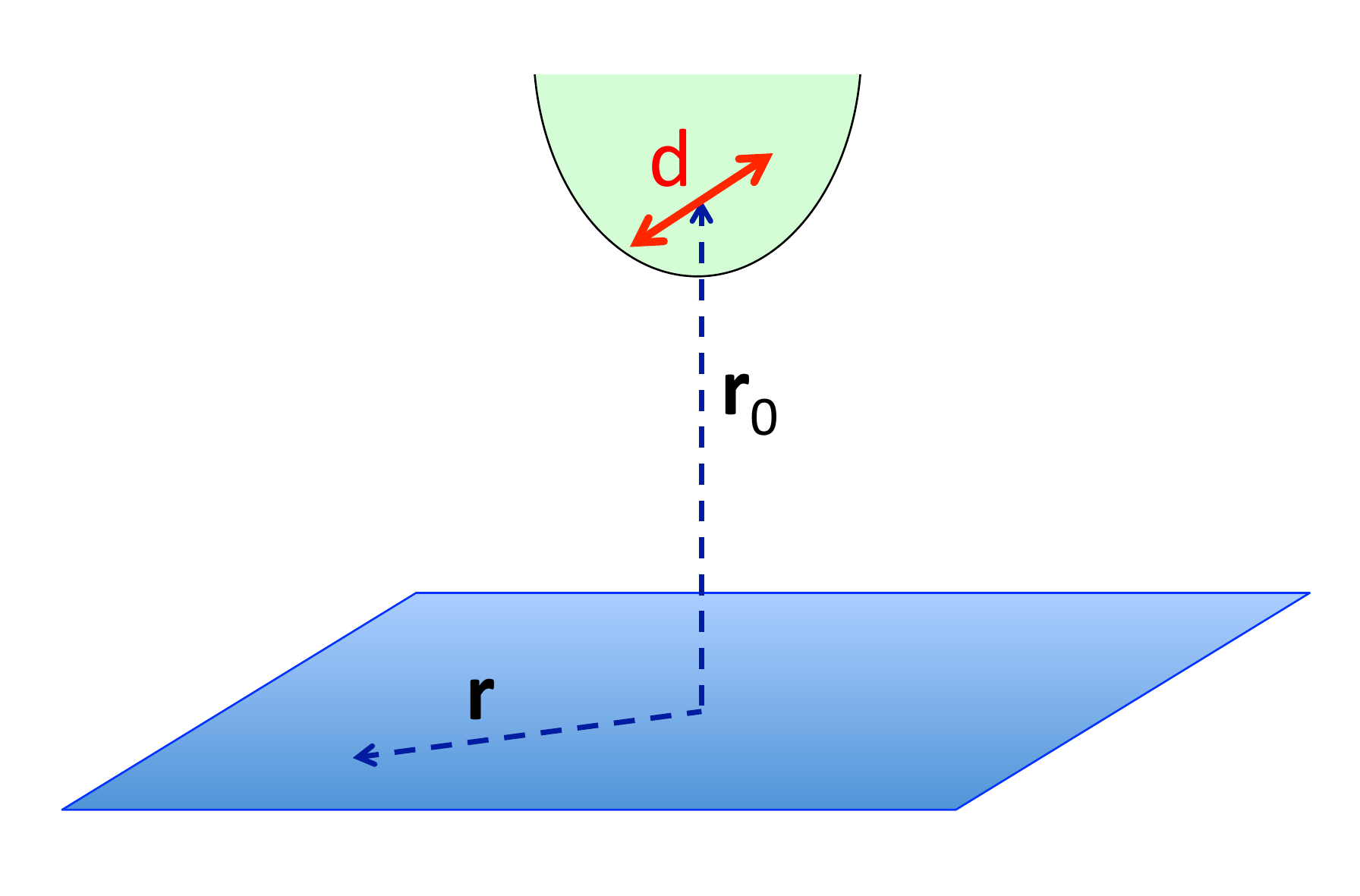}
\caption{Scheme of the SNOM probe.
}
\label{Figure2}
\end{figure}

In the near field approximation, where it is assumed that the distance between the graphene plane and the
dipole is much larger than the dipole size, but it keeps being much smaller than the wavelength of the light emitted by the dipole,
the electric field induced by the dipole is,
\begin{equation}
\phi _{dip} ({\bf r}) = \frac {e ^ {i \omega t}
} {4\pi\epsilon_0\epsilon} \,  \frac {{\bf P} \cdot ({\bf r}- {\bf r}_0)} {|{\bf r} -{\bf r}_0 | ^3 }
\end{equation}
and its two-dimensional  Fourier transform
\begin{equation}
\phi_{dip} ({\bf q})= \frac 1 S \frac {e^{i\omega t}}{2\epsilon_0\epsilon} (-i P_x q_x-i P_y q_y-q P_z) \frac {1}{ q} e ^{i {\bf q}{\boldsymbol \rho} }e ^{-q z_0}
\end{equation}
with $S$ the sample area and the coordinates of the graphene planes are ${\bf r}=({\boldsymbol \rho},0)$.
Using the dielectric constant of the system,                  the total  potential created by the dipole in the graphene layer is
\begin{equation}
\phi _{tot}({\boldsymbol \rho })= \sum _{{\bf q},{\bf G},{\bf G'}} e ^{i ({\bf q}+{\bf G}){\boldsymbol \rho }} \frac {\sqrt{{\bf q}+ {\bf G '}}} {\sqrt{{\bf q}+{\bf G}}} 
e^{- \frac {\delta} 2 (|{\bf q} +{\bf G}|-|\bf q+\bf G'|)} \sum _i \alpha ^ i _ {\bf q} ({\bf G}) \frac {\omega}{\omega -\omega _i ({\bf q})}
\alpha ^i _{{\bf q}}({\bf G}') \phi _{dip} ({\bf q}+ {\bf G}')
\end{equation}
where $\omega _i ({\bf q}) $ and $\alpha _{\bf q} ({\bf G}) $ are the finite frequency eigenvalues and eigenvectors of the matrix
$M$, and the sum in ${\bf q}$ is restricted to the first Brillouin zone of the moir\'e superlattice.
Then, the potential induced by the dipole in the twisted bilayer graphene sheet is 
\begin{equation}
\phi_{ind} ({\boldsymbol \rho})=\phi_{tot} ({\boldsymbol \rho  })-\phi_{dip}({\boldsymbol  \rho})=
\sum _{{\bf q},{\bf G},{\bf G'}} e ^{i ({\bf q}+{\bf G}){\boldsymbol \rho }} \frac {\sqrt{{\bf q}+ {\bf G '}}} {\sqrt{{\bf q}+{\bf G}}} 
e^{- \frac {\delta} 2 (|{\bf q} +{\bf G}|-|\bf q+\bf G'|)} \sum _i \alpha ^ i _ {\bf q} ({\bf G}) \frac {\omega_i ({\bf q})} {\omega -\omega _i ({\bf q})}
\alpha ^i _{{\bf q}}({\bf G}') \phi _{dip} ({\bf q}+ {\bf G}')
\end{equation}
This potential   decreases when  moving away from the bilayer as
\begin{equation}
\phi _{ind} ({\boldsymbol \rho },z)=\sum _{{\bf q}+{\bf G}} \phi _{ind} ({\bf q}+{\bf G}) e ^{ i ({\bf q}+{\bf G}) {\boldsymbol \rho }} e ^{-|{\bf q}+{\bf G}||z|}.
\end{equation}
The induced electric field is
\begin{eqnarray}
E_z ^{ind} ({\boldsymbol \rho},z) &=& \sum _{{\bf q},{\bf G}} {\rm sign} (z) |{\bf q}+{\bf G}| \phi _{ind}({\boldsymbol \rho}, z) \nonumber \\
E_{\nu} ^{ind}({\boldsymbol \rho },z)  &=& -\sum _{{\bf q},{\bf G}} i  ({\bf q} +{\bf G}) \phi _{ind}({\boldsymbol \rho  }, z) \, \, \, \, {\rm for} \, \, \nu=x,y 
\end{eqnarray}

Experimentally the dipole is created by a metallic tip that generally  has an elongate form in the direction perpendicular to the graphene layer, 
and therefore the polarization of the tip is dominant in the $z$-direction. Then, assuming the dipole is orientated in the $z$-direction we get
\begin{eqnarray}
E_z ^{ind} \! ({\boldsymbol \rho },z) &\! \!=\! \! & - e ^{i \omega t} \frac z {|z|} \frac {P_z}{2\epsilon_0\epsilon} \! \! \!\! \sum _{{\bf q},{\bf G},{\bf G}'} \! \!  \! \!F({\bf G},{\bf G}') |{\bf q}\!+\!{\bf G}|
e^{-i({\bf q}+{\bf G}'){\boldsymbol \rho}_0} e^{-|{\bf q}+{\bf G}'|z_0} e^{-|{\bf q}+{\bf G}|z}e ^{i ({\bf q}+{\bf G}){\boldsymbol \rho}} \!\sum _i \! \! \alpha ^ i _ {\bf q} ({\bf G}) \frac {\omega_i ({\bf q})} {\omega \!-\!\omega _i ({\bf q})}
\alpha ^i _{{\bf q}}({\bf G}') 
\nonumber \\
E_{\nu} ^{ind} \!({\boldsymbol \rho },z) & \! \!=\! \!& e ^{i \omega t}\frac {P_z}{2\epsilon_0\epsilon} \! \! \! \!  \sum _{{\bf q},{\bf G},{\bf G}'}  \! \! \! \!   iF({\bf G},{\bf G}')  ({ q}_{\nu}\!+\!{ G}_{\nu})
e^{-i({\bf q}+{\bf G}'){\boldsymbol \rho}_0} e^{-|{\bf q}+{\bf G}'|z_0} e^{-|{\bf q}+{\bf G}|z}  e^{i ({\bf q}+{\bf G}){\boldsymbol \rho}}  \! \sum _i \! \! \alpha ^ i _ {\bf q} ({\bf G}) \frac {\omega_i ({\bf q})} {\omega \! - \! \omega _i ({\bf q})}
\alpha ^i _{{\bf q}}({\bf G}')  \nonumber 
\end{eqnarray}
with 
\begin{equation}
F({\bf G},{\bf G}')=\frac {\sqrt{{\bf q}+ {\bf G '}}} {\sqrt{{\bf q}+{\bf G}}} 
e^{\! -\! \frac {\delta} 2 (|{\bf q} \! + \!{\bf G}|\!-\!|\bf q\!+\!\bf G'|)}\;.
\end{equation}
\begin{figure}[t]
\includegraphics[width=7.5cm]{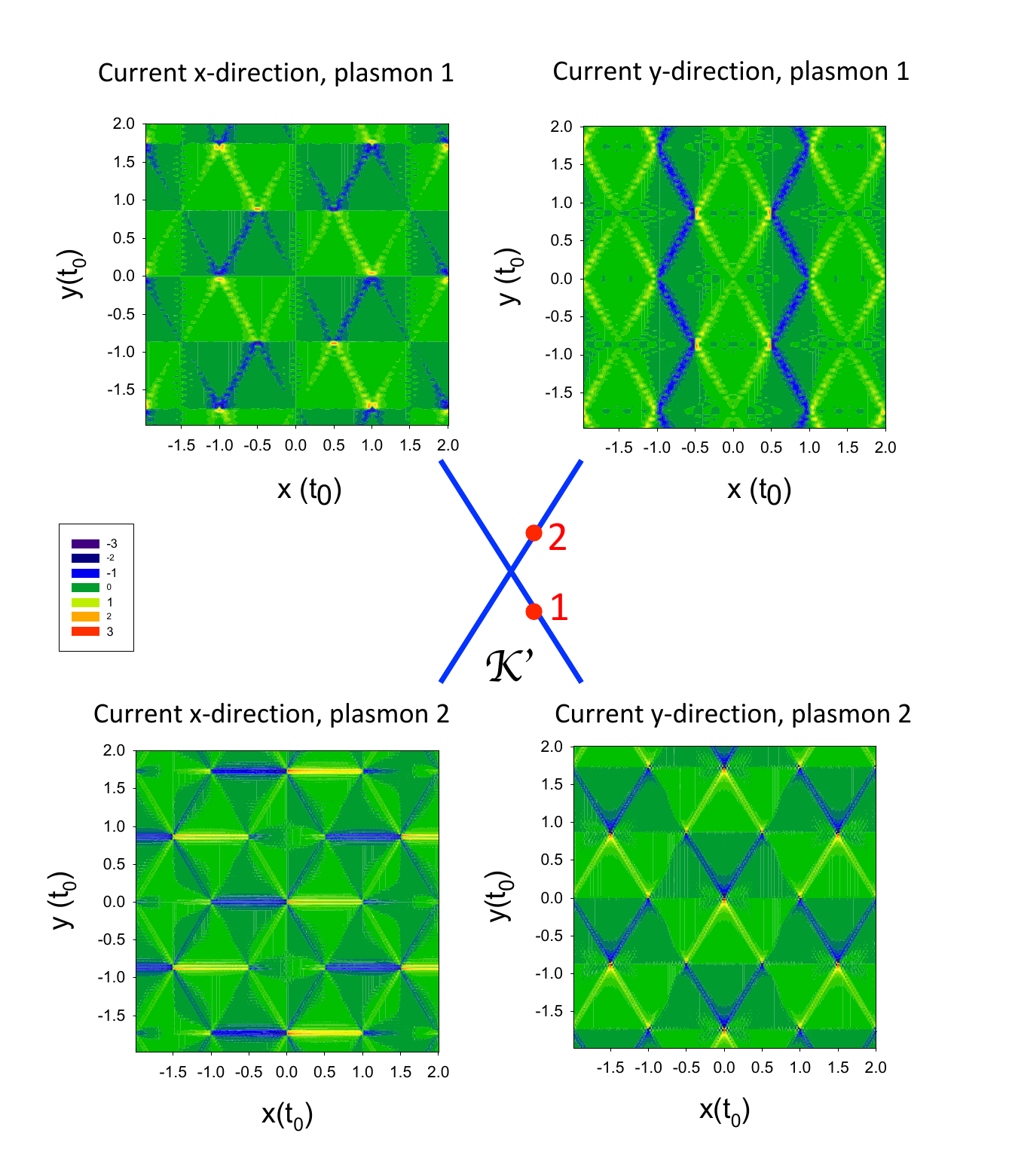}
\caption{Electrical currents corresponding to the two plasmon excitations at the Dirac point. Left panel: Current pattern in $x$ (armchair) direction. Right panel: Current pattern in $y$ (zigzag) direction.}\label{Figure3}
\end{figure}

\section{Plasmonic Dirac cone and gap generation}
\begin{figure}[t]
\includegraphics[width=0.4\columnwidth]{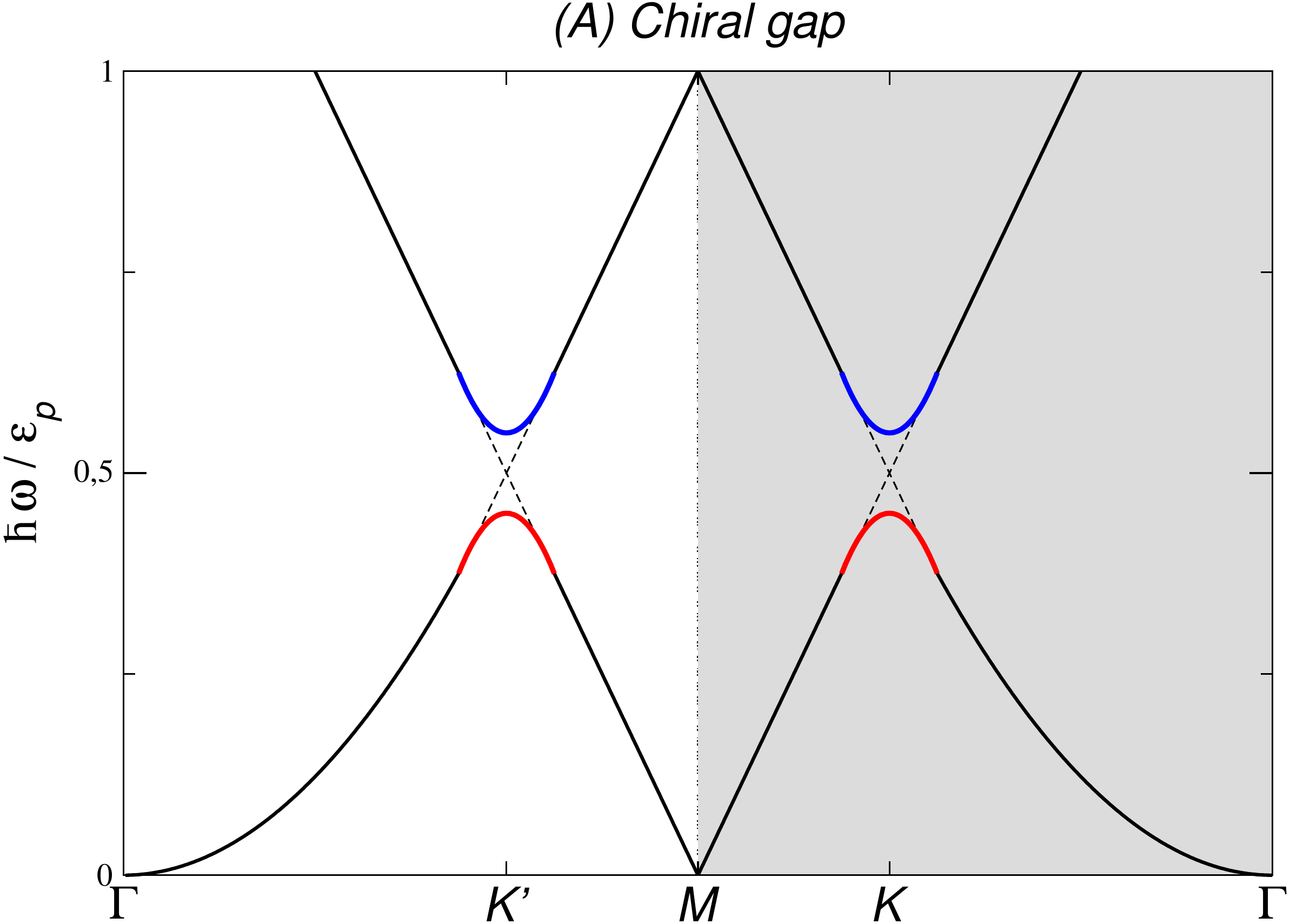}
\includegraphics[width=0.4\columnwidth]{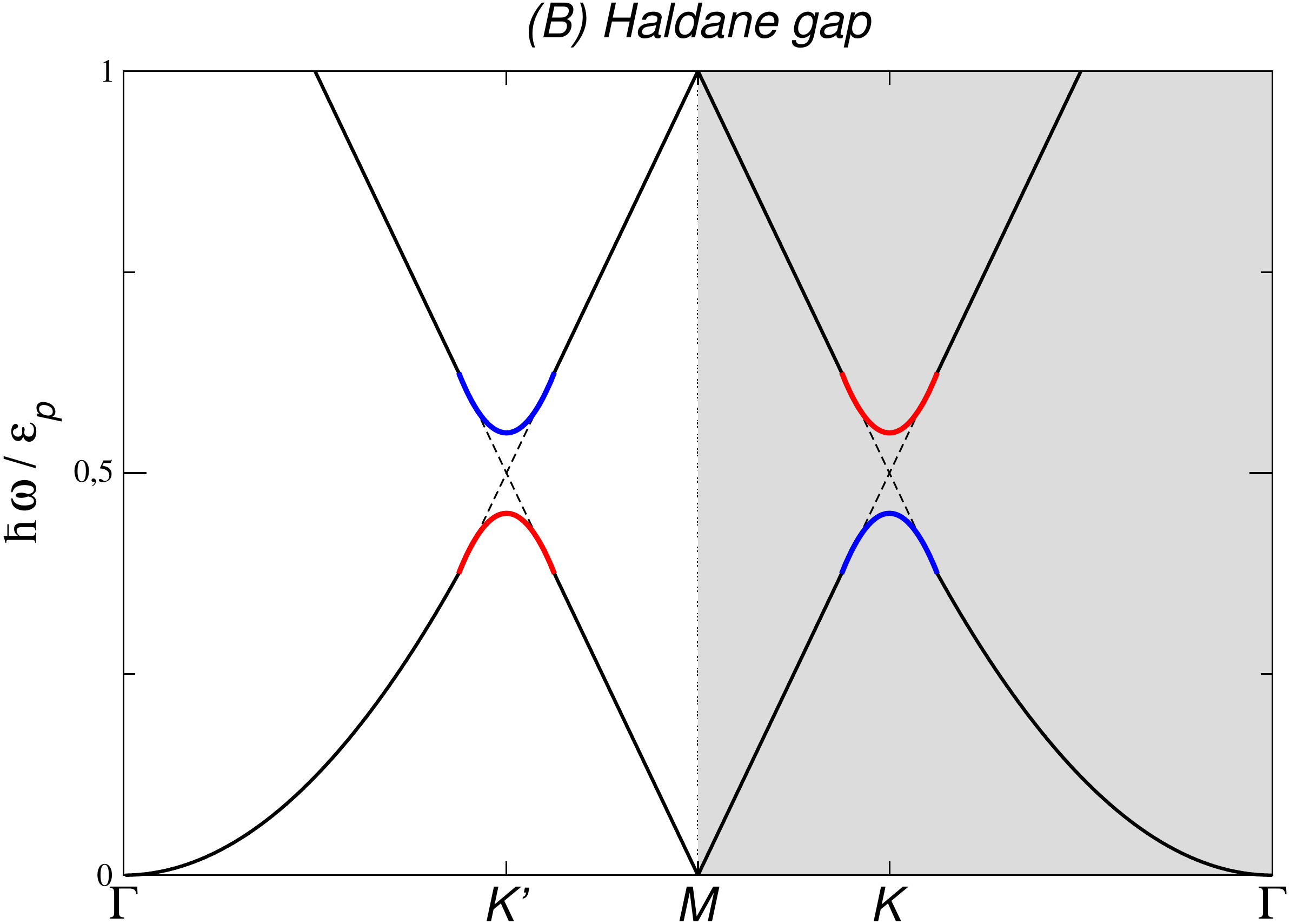}
\caption{Schematic plasmonic dispersion including both graphene valleys indicated by the white and grey shaded regions. (A) The left panel shows a chiral gap opening where both masses have the same sign. (B) The right panel shows a Haldane gap opening where both masses have different sign.}\label{Figure4}
\end{figure}
A general Dirac cone is composed by two orthogonal states that become degenerate at the Dirac point. In the case of graphene, these states correspond to electronic Wannier states localized at the A- and B-lattice, respectively. A gap can be opened up when breaking the corresponding symmetry and in the case of graphene, this would be the inversion symmetry of the A- and B-lattice. 

With regards to the plasmonic (boson) Dirac cone, the orthogonal states are dipole oscillations in the $x$- and $y$-direction. Due to the hexagonal current network, the precise form is given by an armchair- and zigzag-pattern, respectively. This can be seen in Fig. \ref{Figure3}, where the current maps are  shown that build up the Dirac spinor. 

The equivalence of the spinor states can now be broken by the continuous $U(2)$ rotational symmetry. This should favor the possibility of a spontaneous chiral mass generation. Especially for extremely small twist angles with $\theta\lesssim0.1^\circ$, the domain walls are distorted and we expect the plasmonic excitations to be described by a gapped Dirac equation. The equivalence of the spinor states can also be broken by time-reversal symmetry, i.e., a perpendicular magnetic field would favor one orientation of the chiral plasmonic modes.   
\begin{figure}[t]
\includegraphics[width=0.4\columnwidth]{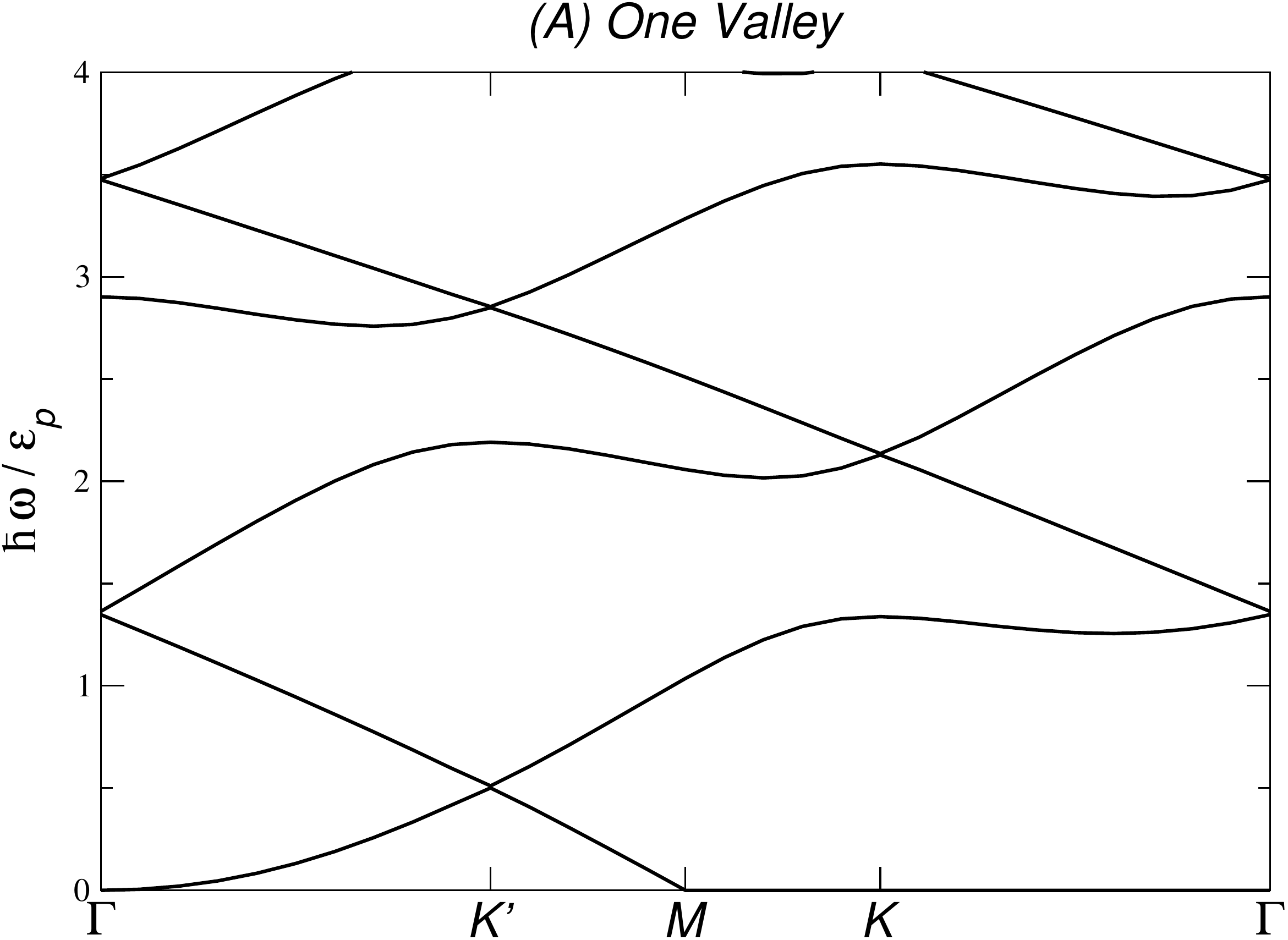}
\includegraphics[width=0.4\columnwidth]{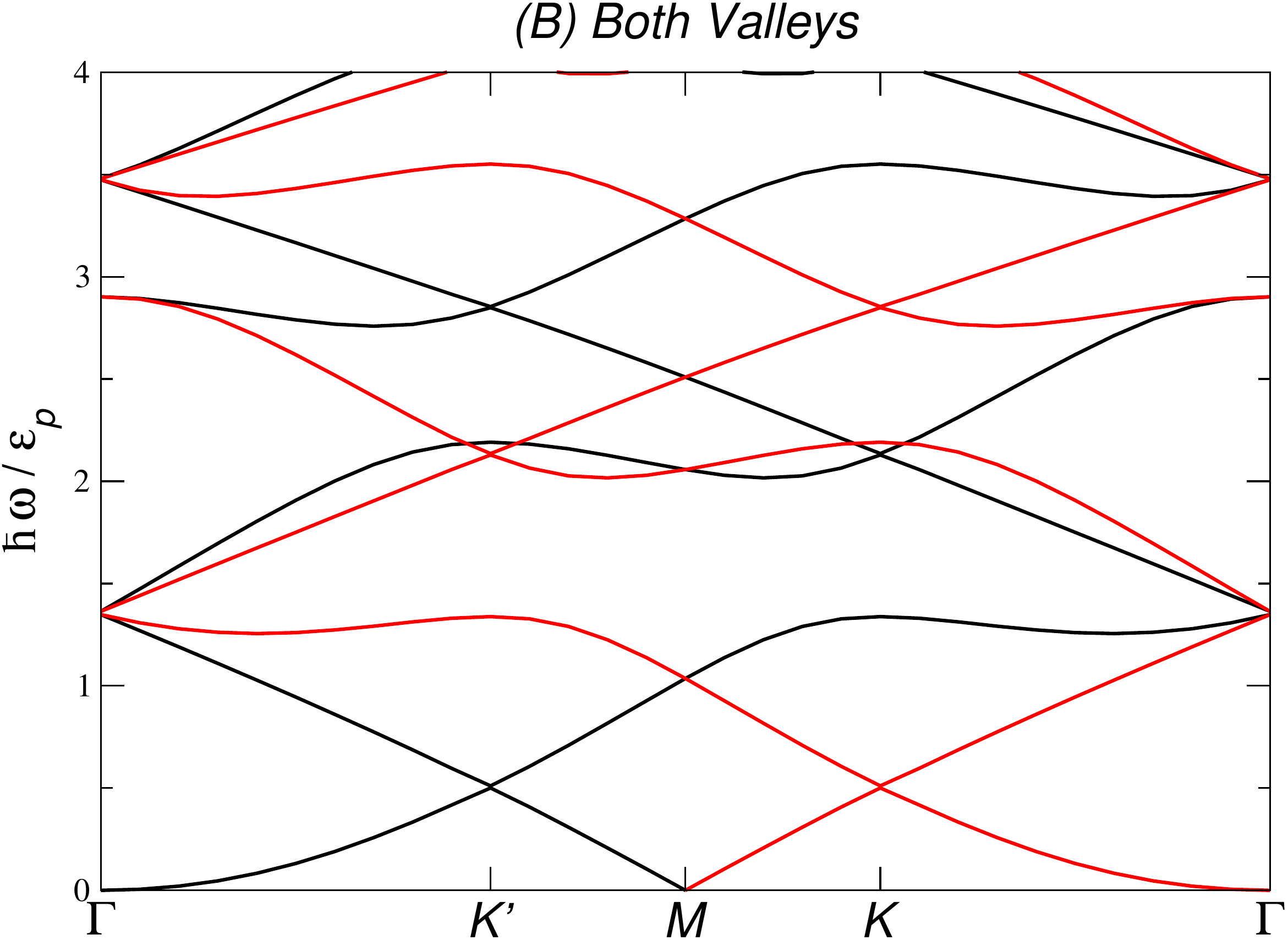}
\caption{(A) Plasmonic dispersion of one valley, also displaying the alternating Dirac cones. (B) Plasmonic dispersion including both valleys indicated as black and red curves.}\label{Figure5}
\end{figure}

The plasmonic Dirac cones of the two graphene valleys are located on different regions of the moir\'e Brillouin zone. This opens up the possibility of a Haldane gap generation where the two gaps have different signs. The resulting topologically protected plasmonic edge currents along the zone boundary might be truly protected from backscattering as well as from decaying into particle-hole excitations as they are composed of topologically protected electronic states. In Fig. \ref{Figure4}, we contrast the two different gap openings schematically. On the left panel, a chiral gap opening is shown where both masses have the same sign, whereas on the right panel, the Haldane gap is defined with different mass signs for the two valleys. Let us also mention the possibility of Dirac velocity renormalization due to vertex corrections or retardation effects.

Finally, we want to mention that Dirac cones also appear at higher energies. We observe an alternating hierarchy where the Dirac cones of one valley change the two separate sector of the Brillouin zone, see Fig. \ref{Figure5}. They are connected by Dirac cones that appear at the $\Gamma$-point. The energy of the additional Dirac cones depends stronger on the cut-off energy $\delta$ given a fixed maximal reciprocal lattice vector $G_{max}$. Still, together with the van Hove singularities they might be observable in samples with large lattice constant $t_0$.

\end{widetext}

%


\vspace{0.5truecm}
\end{document}